\documentclass[twocolumn,times,amsmath]{aastex702}
\usepackage[T1]{fontenc}
\usepackage[utf8]{inputenc}
\usepackage{booktabs}
\usepackage{etoolbox}
\newcommand{\mdot}{\dot M}

\newcommand{\figfile}[2][]{\includegraphics[width=\linewidth,#1]{#2}}

\makeatletter
\renewcommand{\frontmatter@title@above}{}
\makeatother
\shorttitle{Accretion heating and carbon instability}
\shortauthors{Hou, Aoyama, Dohi}
\journalinfo{}
\ifdefined\pdfinfoomitdate\pdfinfoomitdate=1\fi
\hypersetup{
  pdftitle={Constraining Accretion Heating in Neutron Stars with Carbon Instability and Quiescent Cooling},
  pdfauthor={Chengze Hou, Amira Aoyama, Akira Dohi},
  hidelinks
}

\makeatletter
\patchcmd{\frontmatter@title@produce}
  {\hskip-11ptEmail: \@@email}
  {\hskip-11pt\parbox[t]{\dimexpr(\textwidth-\columnsep)/2\relax}{\raggedright Email: \@@email}}
  {}{\PackageError{aastex-layout}{Email patch failed}{}}
\newcommand{\firstpagetextbottom}{\ifnum\c@page=1 \vskip\z@\@plus.0001fil\fi}
\AtBeginDocument{\flushbottom\@booleanfalse\raggedcolumn@sw\let\@textbottom\firstpagetextbottom}
\patchcmd{\section}{{9pt plus 1pt minus 1pt}{4pt}}{{9pt plus 1pt minus 1pt}{0.01pt}}
  {}{\PackageError{aastex-layout}{Section spacing patch failed}{}}
\patchcmd{\subsection}{{9pt plus 1pt minus 1pt}{4pt}}{{9pt plus 1pt minus 1pt}{0.01pt}}
  {}{\PackageError{aastex-layout}{Subsection spacing patch failed}{}}
\patchcmd{\@acknowledgments}{\vskip 5.8mm plus 1mm minus 1mm}{\vskip0pt}
  {}{\PackageError{aastex-layout}{Acknowledgments spacing patch failed}{}}
\patchcmd{\HyOrg@appendix}{\vskip18pt}{\vskip6pt}
  {}{\PackageError{aastex-layout}{Appendix spacing patch failed}{}}
\patchcmd{\HyOrg@appendix}{\vskip18pt}{\vskip6pt}
  {}{\PackageError{aastex-layout}{Appendix spacing patch failed}{}}
\makeatother
\begin{document}
\setlength{\parskip}{0pt plus 0.2pt}
\setlength{\baselineskip}{12.5pt plus 0.2pt}
\setlength{\textfloatsep}{8pt plus 2pt minus 2pt}
\renewcommand{\textfraction}{0.15}
\renewcommand{\dbltopfraction}{0.85}
\renewcommand{\dblfloatpagefraction}{0.5}
\renewcommand{\floatpagefraction}{0.5}
\title{Constraining Accretion Heating in Neutron Stars with Carbon Instability and Quiescent Cooling}

\author{Chengze Hou}
\affiliation{Department of Earth and Planetary Science, University of California, Berkeley, California 94720, USA}
\email[show]{h\_chengze@berkeley.edu}

\author[0009-0008-3926-363X]{Amira Aoyama} 
\affiliation{Department of Physics, Tokyo University of Science, 1-3 Kagurazaka, Shinjuku, Tokyo 162-8601, Japan}
\affiliation{RIKEN Pioneering Research Institute (PRI), 2-1 Hirosawa, Wako, Saitama 351-0198, Japan}
\email[show]{amira.aoyama@riken.jp}

 \author[0000-0001-8726-5762]{Akira~Dohi}
 \affiliation{Interdisciplinary Theoretical and Mathematical Sciences (iTHEMS), RIKEN, Wako, Saitama 351-0198, Japan}
 \email[show]{akira.dohi@riken.jp}

\begin{abstract}
In accreting neutron stars, different heating depths can produce similar quiescent cooling when the heating strength and core temperature are adjusted. For KS~1731$-$260, we compare thermal histories with accretion heating either in a shallower region that overlaps the carbon fuel layer or in a deeper region below it. Without the carbon instability condition, the two histories fit ten quiescent \textit{Chandra} spectra nearly equally well. When local carbon instability is required at the observed superburst epoch, the deeper region needs stronger accretion heating than the shallower region. After the thermal and spectral parameters are refitted, the deeper history retains more heat beneath the fuel and produces excess early quiescent emission. A larger fitted absorbing column partly attenuates this excess but makes the late spectra too faint. The fuel temperatures at the superburst epoch are similar in the two histories, so the same prescribed energy release produces nearly identical superburst decays. Including carbon burial, consumption and nuclear heating reduces the accretion heating needed for rapid burning to begin near the observed superburst; shallower heating still fits the quiescent spectra better. This preference persists across the tested fuel depths, fusion rates, distances and absorption treatments. In synthetic examples, carbon in older material allows deeper heating to match the cooling of a shallower-heating reference. Inferring the heating location therefore requires a fuel distribution and thermal history consistent with ignition, burst emission and quiescent cooling.
\end{abstract}
\section{Introduction}
\label{sec:introduction}

Neutron stars undergo deep crustal heating during accretion. Newly accreted material compresses the crust to higher densities, triggering electron captures and other nuclear reactions that deposit heat locally \citep{haensel2008}. After accretion stops, thermal X-ray observations trace the cooling of the heated crust \citep{2018JPhG...45i3001M}. Models that include deep crustal heating often need additional shallow heating of uncertain origin to reproduce the observed cooling \citep{brown2009,jain2025}. Different depths for this additional heating can produce similar cooling curves when the heating strength and core temperature are adjusted.

A superburst observed during accretion can help distinguish heating models for the subsequent quiescent cooling. Superbursts last for several hours and are attributed to unstable carbon burning in the neutron star ocean \citep{2001ApJ...559L.127C}. The temperature required for carbon ignition depends on the fuel depth and carbon abundance. Previous studies have estimated the heating needed to reach this temperature \citep{deibel2016,meisel2024}. After carbon burning heats the fuel, the burst fades as the heated layer cools. Fitting this decay constrains the depth of the heated layer and the energy deposited per unit mass \citep{cumming2006,keek2015}. We take the bottom of this layer as the fuel depth for the ignition calculation. For an assumed nuclear energy yield and complete carbon consumption, the inferred energy release per unit mass also gives the carbon abundance before the burst. These decay constraints let us test whether heating at different depths can bring carbon to instability at the observed superburst epoch while reproducing quiescent cooling.

We examine this question using the superburst and quiescent cooling of KS~1731$-$260 (hereafter KS). The superburst was detected in 1996 during prolonged accretion \citep{kuulkers2002}. The source was observed in quiescence in 2001, and its cooling was monitored for more than a decade \citep{cackett2006,merritt2016}. \citet{merritt2016} fitted the cooling by adjusting the heating strength within a fixed deposition region; the resulting curves were relatively insensitive to the precise boundaries of that region. \citet{meisel2024} found that those heating estimates were difficult to reconcile with carbon ignition at the superburst depth inferred from the decay. Meisel suggested that fixing the heating depth could contribute to the discrepancy.

We test this possibility by comparing two heating regions in KS: one overlaps the carbon fuel layer and the other lies below it. Accretion variability changes the temperature profile at the start of quiescence \citep{ootes2016}, while the ash composition alters heat transport and the heating inferred from cooling \citep{jain2025}. To isolate the effect of heating location, we use a common accretion history and transport prescription for both regions. We fit the corresponding thermal histories to ten individual \textit{Chandra} spectra, both with and without imposing local carbon instability at the superburst epoch. Requiring local carbon instability favors heating that overlaps the fuel layer. We also examine how carbon burial and burning affect this inference, including the possibility of fuel surviving from earlier accretion.

This paper is organized as follows. Section~\ref{sec:methods} describes the observations, thermal models and fuel calculations. Section~\ref{sec:results} presents the heating comparison, burst decay, carbon evolution and synthetic alternatives. Section~\ref{sec:discussion} relates these results to fuel accumulation, burning and heat transport, and considers their application to other sources. Section~\ref{sec:conclusions} summarizes the implications for inferring the heating location.

\begin{figure*}[t]
\centering
\figfile[width=0.87\linewidth]{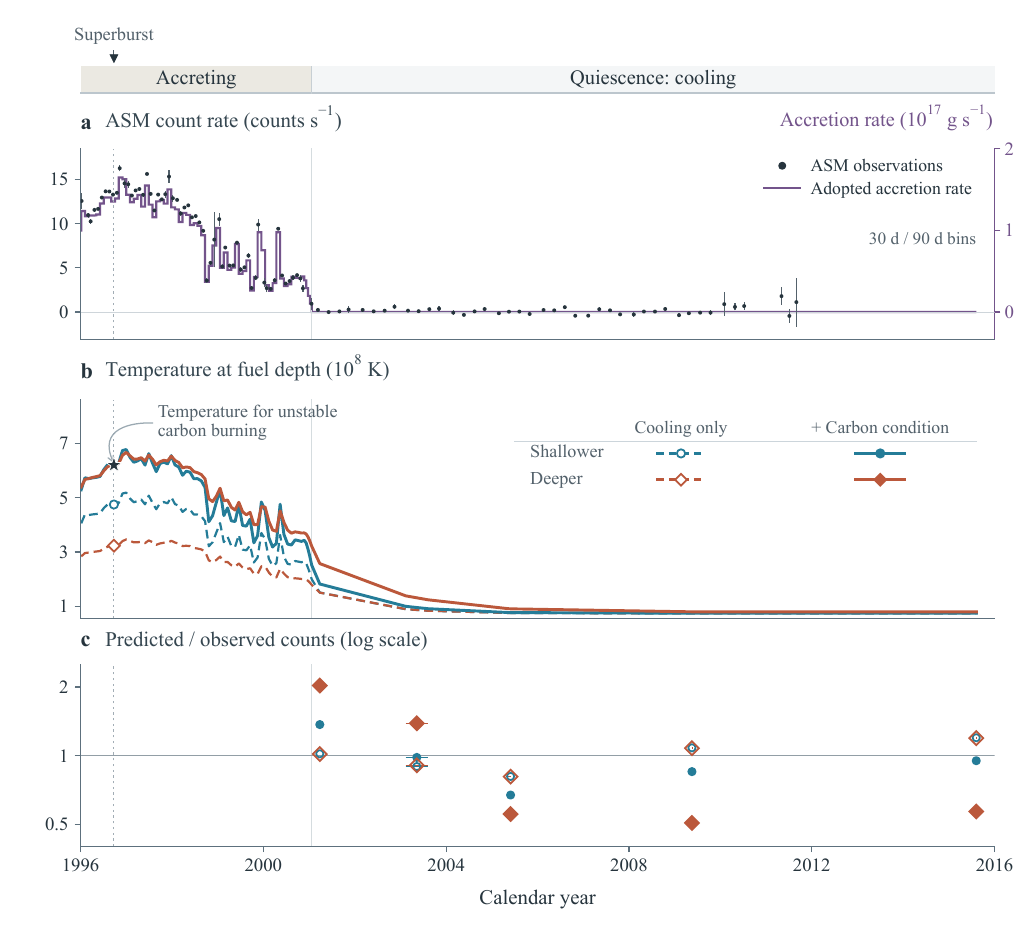}
\caption{Thermal histories and cooling observations of KS~1731$-$260 for $y_*=6.3\times10^{11}$~g\,cm$^{-2}$ and $X_{\rm C}=0.20$. The upper strip separates the modeled accretion and cooling intervals. (a) RXTE ASM count rates in the 1.5--12~keV band, averaged in 30~d bins before the adopted end of accretion and 90~d bins afterward, with propagated measurement errors. The purple curve shows the adopted accretion rate at 4.5~kpc on the right axis (Appendix~\ref{app:history}). (b) Background temperature at the fuel depth; the star marks the calculated temperature for local carbon instability at the observed superburst epoch. (c) Ratios of predicted to observed source-region counts for the five observing years, shown on a logarithmic axis. Each ratio uses counts summed over one year; predictions include the scaled measured background. In (b,c), dashed lines and open symbols show fits without the carbon instability condition; solid lines and filled symbols show fits requiring local carbon instability. Horizontal spans in (c) show the range of exposure midpoints within each observing year.}
\label{fig:timeline}
\end{figure*}

\section{Observational constraints and heating scenarios}
\label{sec:methods}

\subsection{Observations and Spectral Fitting}
\label{sec:observations}

We use the published fluxes of the 1996 superburst in 18 time intervals \citep{kuulkers2002} and ten quiescent Chandra spectra from 2001--2015 \citep{cackett2006,merritt2016}. Because the transition to quiescence was not continuously observed, we measure cooling times from the last active detection in 2001 and adopt that epoch as the end of accretion in the thermal histories. We use Chandra alone for the spectral fit to maintain consistent calibration and extraction procedures across all exposures (Appendix~\ref{app:spectra}).

Both heating scenarios share the same mass accretion rate $\mdot(t)$, where $t$ is observer time. We reconstruct this rate from RXTE ASM count rates and X-ray colors measured during the outburst and normalize it to the persistent flux before the superburst (Figure~\ref{fig:timeline}a). The adopted rate is constant before ASM monitoring and interpolated across the final monitoring gap (Appendix~\ref{app:history}). The reference KS calculations use stellar mass $M=1.4\,M_\odot$, radius $R=10$~km and distance $d=4.5$~kpc, as in \citet{cumming2006}.

For each trial thermal history, we evaluate the local surface temperature at the midpoint of each of the ten quiescent exposures. We calculate the gravitationally redshifted emission at these temperatures with NSA \citep{zavlin1996} in XSPEC \citep{arnaud1996}. The model assumes a nonmagnetic hydrogen atmosphere covering the entire stellar surface. Only the first exposure includes an additional power law, whose amplitude is free. Both components undergo interstellar photoelectric absorption, described by PHABS with absorbing hydrogen column $N_{\rm H}$. We then fold the absorbed spectrum through the effective area and energy redistribution matrix for each exposure to predict detector counts.

We fit the source and background spectra from all ten observations together for each trial thermal history. The fit uses the Poisson spectral statistic $W$ over 540 energy groups; lower values indicate closer agreement. With the ten atmosphere temperatures fixed by each trial thermal history, we minimize $W$ over two parameters: a single $N_{\rm H}$ shared by all ten exposures and the first-exposure power-law amplitude (Appendix~\ref{app:spectra}).

\begin{table*}[t]
\centering
\caption{KS cooling fits at $y_*=6.3\times10^{11}$~g\,cm$^{-2}$.}
\label{tab:ksfits}
\begin{tabular}{lrrrr}
\toprule
 & \multicolumn{2}{c}{Shallower Region} & \multicolumn{2}{c}{Deeper Region}\\
\cmidrule(lr){2-3}\cmidrule(lr){4-5}
Quantity & Cooling & $+$ Carbon Instability & Cooling & $+$ Carbon Instability\\
\midrule
$q_a$ (MeV per nucleon) & 1.39 & 2.19 & 1.34 & 4.00\\
$T_{\rm c}$ ($10^7$ K) & 7.73 & 7.77 & 7.72 & 8.32\\
$N_{\rm H}$ ($10^{22}$ cm$^{-2}$) & 1.35 & 1.59 & 1.35 & 2.48\\
Spectral statistic $W$ & 267 & 306 & 266 & 618\\
\bottomrule
\end{tabular}
\begingroup
\patchcmd{\tablecomments}{{\sc Note}---}{}{}{\PackageError{aastex-layout}{Table-note label patch failed}{}}
\tablecomments{In the constrained models, the fuel has $X_{\rm C}=0.20$ and is locally unstable at the observed superburst epoch.}
\endgroup
\end{table*}

\subsection{Heating Locations and the Carbon Instability Requirement}
\label{sec:heatingcriteria}

For each heating region, we first fit the quiescent spectra without the carbon instability condition (the cooling-only fit). We then refit while requiring local carbon instability at the observed superburst epoch. The fuel is specified by its column depth $y_*$ and carbon mass fraction $X_{\rm C}$. At the reference distance, the superburst decay is fitted with layer cooling models that assume uniform energy release per unit mass. From these preliminary fits, we select three representative values of $y_*$ spanning $6.3\times10^{11}$--$10^{12}$~g\,cm$^{-2}$ (Appendix~\ref{app:history}). For these comparisons, we identify the bottom of the layer heated by the burst with the depth at which we impose the carbon instability condition.

The shallower heating region includes all three fuel depths; the deeper region lies below them, so heat deposited there reaches the fuel by outward conduction. With local pressure $P$ expressed in units of dyn\,cm$^{-2}$, the deposition intervals are $\log_{10}P=26$--27 and 27--28, respectively. In the thin crust approximation, the column coordinate is $y=P/g$, where $g$ is the neutron star surface gravity. The heating strength in region $a$ is $q_a$, in MeV per accreted nucleon. The local core temperature $T_{\rm c}$ remains constant within each history.

We solve the radial heat transport equation during accretion and quiescence with dStar \citep{brown2015} on a fixed hydrostatic structure to obtain the local temperature $T$. For the KS thermal histories, we use the prescribed accreted crust of \citet{haensel1990}, denoted HZ90, and a static envelope with helium above iron \citep{brown2009}; Appendix~\ref{app:implementation} gives the transport assumptions. The adjustable source replaces standard crustal heating within its deposition interval and scales with $\mdot(t)$. We express the conductive heat flow as the redshifted luminosity $L^\infty$, positive outward.

\begingroup
\setlength{\baselineskip}{12.5pt}
\setlength{\abovedisplayskip}{4pt plus 1pt minus 1pt}
\setlength{\belowdisplayskip}{4pt plus 1pt minus 1pt}
\setlength{\abovedisplayshortskip}{4pt plus 1pt minus 1pt}
\setlength{\belowdisplayshortskip}{4pt plus 1pt minus 1pt}
We evaluate local carbon instability using thermal histories evolved without carbon production or burning. The carbon heating rate $\epsilon_{\rm C}$ and estimated conductive cooling rate $\epsilon_{\rm cool}$ are both expressed per unit mass in erg\,g$^{-1}$\,s$^{-1}$. At the fuel depth and the observed superburst epoch, the carbon heating rate must increase with temperature at least as rapidly as the conductive cooling rate at fixed pressure~\citep{1981ApJ...247..267F}:
\begin{equation}
 \left(\frac{\partial\epsilon_{\rm C}}{\partial T}\right)_P
 \geq
 \left(\frac{\partial\epsilon_{\rm cool}}{\partial T}\right)_P.
 \label{eq:ignition}
\end{equation}
\endgroup
The local temperature at equality is $T_{\rm crit}$. In the reference calculations, we adopt the carbon fusion rate of \citet[hereafter CF88]{cf1988}, with Coulomb screening from \citet{chugunov2007}, for a prescribed mixture with carbon mass fraction $X_{\rm C}=0.20$ and iron mass fraction 0.80. Conductive cooling is represented by a one-zone loss estimate \citep{potekhin2012,deibel2016,meisel2024}; Appendix~\ref{app:material} gives the microphysics.

In both the cooling-only and constrained fits, we readjust $q_a$, $T_{\rm c}$ and the spectral parameters at each fuel depth (Appendix~\ref{app:search}). The carbon instability condition is imposed as a constraint on the allowed thermal histories. For each heating region, $\Delta W$ is the increase in $W$ relative to the cooling-only fit and measures the loss of spectral agreement. We assess the absolute goodness of fit with count simulations that hold each selected thermal history fixed (Appendix~\ref{app:spectra}).

For the same measured fluxes, a larger distance implies a higher accretion rate and quiescent luminosity. To test the distance dependence of the inferred heating location, we repeat the comparison at 7~kpc, the distance used by \citet{cackett2006} and \citet{merritt2016}, keeping the mass, radius and smallest fuel depth fixed.

The carbon fusion rate affects the temperature required for ignition \citep{2009ApJ...702..660C,2022ApJ...937..124D}. We also evaluate the carbon instability condition using the faster fusion rate inferred by \citet[hereafter T18]{tumino2018}. At the smallest and largest fuel depths, both rates are evaluated with $X_{\rm C}=0.20$ and the same screening prescription. The middle-depth comparison uses the common carbon abundance fitted to the decay as described in Section~\ref{sec:decaymethod}.

\subsection{Fuel Constraints and Evolution}
\label{sec:decaymethod}

The carbon abundance affects both the instability temperature and the energy released by burning. For complete burning to iron-group ashes at a fixed energy yield, a uniform carbon mass fraction $X_{\rm C}$ corresponds to a uniform energy release per unit mass. At $X_{\rm C}=0.20$, the release is about $2\times10^{17}$~erg\,g$^{-1}$, comparable to the energy per unit mass inferred from superburst decay models~\citep{cumming2006,keek2015}. For each trial $X_{\rm C}$, we deposit the corresponding energy instantaneously throughout the layers shallower than $y_*$. The predicted burst flux is the excess above the flux from an otherwise identical evolution without this deposition. At each fuel depth, the fits with local carbon instability provide one thermal history for each heating region. Fitting both histories to the burst fluxes in all 18 time intervals gives a common $X_{\rm C}$ and time of energy release (the model onset). The thermal histories are then refitted with the carbon instability condition evaluated at this abundance. We recompute both decays on the refitted histories (Appendix~\ref{app:burstdecay}).

To examine how the burst energy distribution affects the depth inferred from the decay, we allow the bottom column of the heated layer, $y_{\rm h}$, to vary. We compare two distributions: uniform specific energy release and release increasing as $(y/y_{\rm h})^{1/4}$~\citep{keek2015}. For each distribution, we fit $y_{\rm h}$, the mean energy deposited per gram and the onset on the shallower and deeper CF88 thermal histories selected at the middle fuel depth. The thermal backgrounds, fuel depths and carbon abundances remain fixed during the decay fits.

Burial and burning can change the assumed fuel distribution. For KS, we first calculate carbon consumption using the fitted temperature histories, then include nuclear heating in the thermal evolution. For both heating regions and fusion rates, we adjust accretion heating so that the first rapid temperature rise occurs within about a month of the observed superburst epoch. In all four histories, the core temperature remains fixed at the value from the corresponding prescribed-fuel fit. We follow carbon supply, consumption and nuclear heating through the burning episode, the remaining accretion and all ten quiescent exposures. Heat transport is conductive throughout; convection is omitted. For each history, we refit the common absorbing column and first-exposure power-law amplitude to the quiescent spectra. To compare burst shapes, we fit a time offset between each calculated light curve and the observed decay (Appendix~\ref{app:coupledburning}).

\looseness=1 Carbon from earlier accretion episodes may survive below the newly accreted layer. In a separate synthetic comparison, we test whether a history with deeper heating can reproduce the cooling of a shallower-heating reference if carbon survives in this older material. Both histories must satisfy the carbon instability condition. All alternatives share the same constant accretion rate and burning envelope \citep{nava2025}. The column $Y_0$ is the mass accreted from the start of the episode to the chosen superburst epoch divided by the stellar surface area. Material below this column therefore predates the episode. At that epoch, we prescribe $X_{\rm C}=0.20$ at each candidate depth and find the shallowest depth satisfying Equation~\ref{eq:ignition}. The cooling comparison uses each alternative's observer-frame effective temperature at seven epochs after accretion ends. The reported difference is the largest absolute deviation from the reference, in eV (Appendix~\ref{app:implementation}).

\begin{figure*}[t]
\centering
\figfile[width=1.0\linewidth]{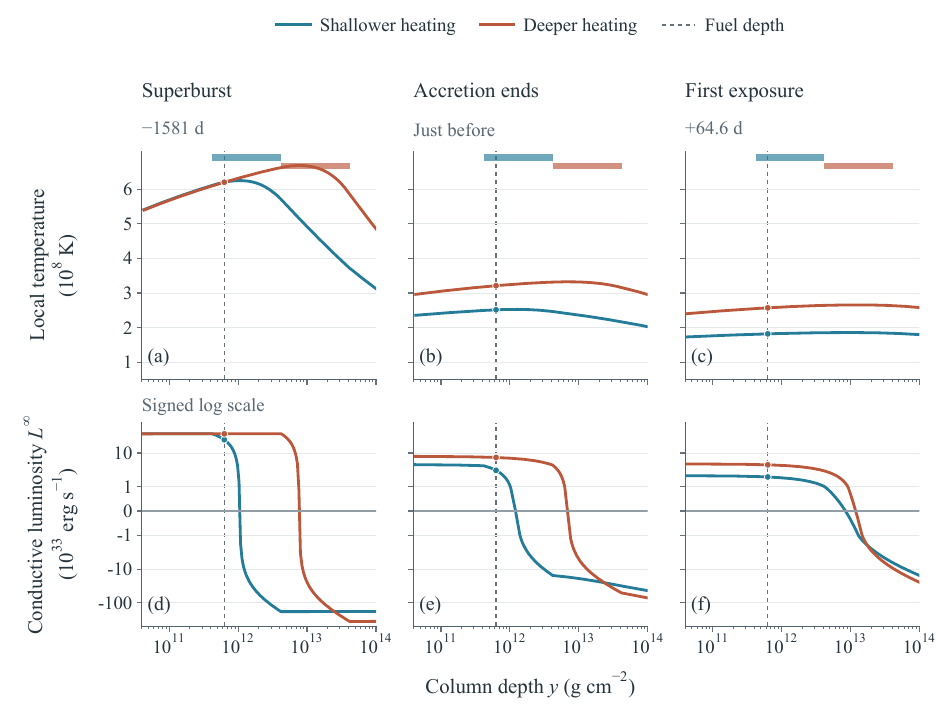}
\caption{Temperature and conductive luminosity profiles for the two histories fitted with local carbon instability at $y_*=6.3\times10^{11}$~g\,cm$^{-2}$ and $X_{\rm C}=0.20$. Times are relative to the adopted end of accretion. Dashed vertical lines mark the fuel depth, and dots mark the corresponding profile values. Colored bars indicate the deposition intervals used during accretion; accretion heating has ended in the last column.}
\label{fig:thermal}
\end{figure*}

\section{Results}
\label{sec:results}

\subsection{The Carbon Instability Condition Favors Shallower Heating}
\label{sec:ksfits}

Without the carbon instability condition, histories with shallower and deeper heating fit the quiescent spectra nearly equally well (Table~\ref{tab:ksfits}). Their fuel temperatures differ during accretion, but both lie below the instability temperature $T_{\rm crit}$ at the observed superburst epoch (Figure~\ref{fig:timeline}b).

\looseness=1 At the smallest fuel depth, bringing carbon to instability requires nearly twice as much heat per accreted nucleon from the deeper source as from the shallower one (Table~\ref{tab:ksfits}). After refitting, the shallower and deeper histories predict about 1.4 and 2 times the observed count total in the first exposure, respectively. The deeper history also underpredicts the later counts (Figure~\ref{fig:timeline}c). The carbon instability condition worsens the fit to the full spectral sequence more for deeper heating ($\Delta W\simeq352$) than for shallower heating ($\Delta W\simeq39$).

\begin{figure*}[t]
\centering
\figfile[width=1.0\linewidth]{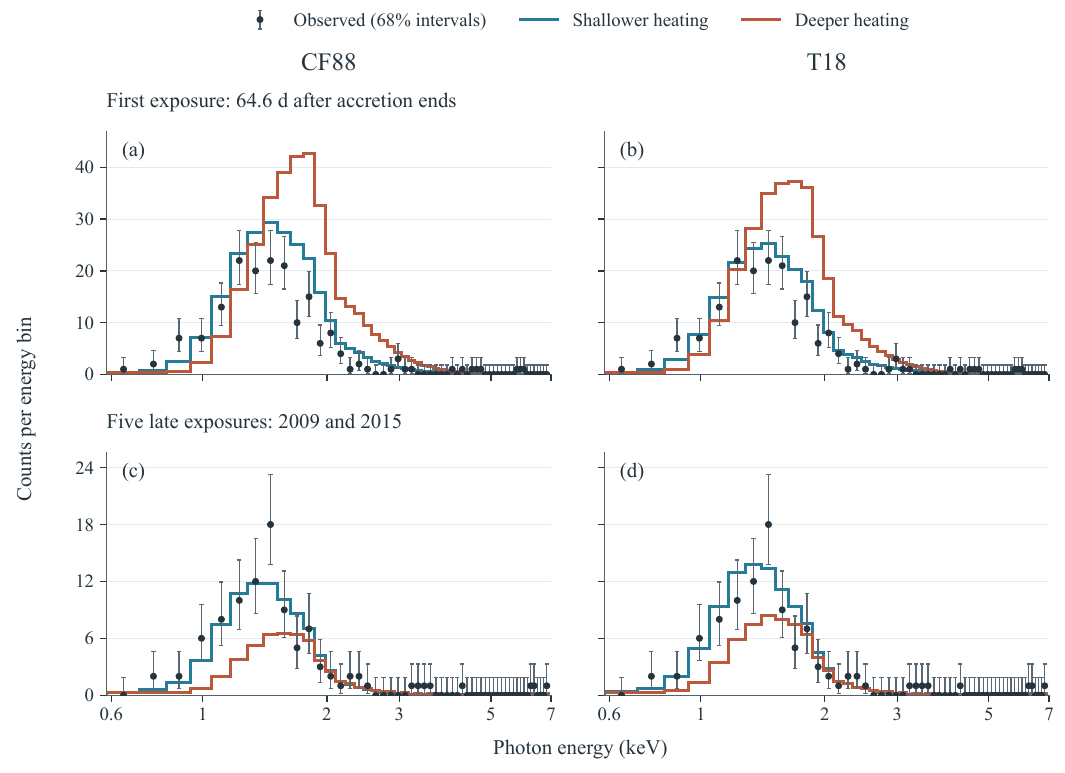}
\caption{Quiescent spectra fitted with local carbon instability at $y_*=6.3\times10^{11}$~g\,cm$^{-2}$ and $X_{\rm C}=0.20$, using CF88 (left) and T18 (right). Panels (a,b) show the first exposure; (c,d) sum the five exposures in 2009 and 2015. Curves show an NSA atmosphere, with an additional power law in the first exposure. The combined emission is absorbed and folded through the instrumental responses, then added to the scaled measured background. Points and error bars show source-region counts and central 68\% Poisson intervals; the same observations appear in both columns.}
\label{fig:spectra}
\end{figure*}

\looseness=1 For the selected CF88 histories at the smallest fuel depth, the spectral disagreement after imposing carbon instability is larger than expected from count fluctuations. In simulations conditional on each thermal history and its fitted background, none of 300 realizations has $W$ as large as observed. For the cooling-only fits, about one in ten does (Appendix~\ref{app:spectra}). 

At 7~kpc, histories with either heating region still fit the quiescent spectra nearly equally well without the carbon instability condition. Imposing this condition at the same fuel depth degrades the spectral fits less than at 4.5~kpc. The shallower history's first-exposure excess falls by more than half, whereas the deeper history retains a substantial early excess and late deficit. The preference for shallower deposition therefore persists at this distance (Appendix~\ref{app:search}).

\subsection{Stored Heat and Absorption Explain the Spectral Residuals}
\label{sec:transport}
\label{sec:spectra}

At the reference distance, the two histories fitted with carbon instability have surface thermal luminosities that agree to within 1\% at the superburst epoch (Figure~\ref{fig:thermal}a,d). Beneath the fuel layer, the temperature profile for deeper heating reaches a higher maximum at greater depth.

Just before accretion ends, the deeper history has both a hotter interior and a larger outward heat flow through the fuel layer (Figure~\ref{fig:thermal}b,e). At the first exposure, its interior remains hotter (Figure~\ref{fig:thermal}c), and both the outward heat flow and the surface thermal luminosity are about twice those of the shallower history (Figure~\ref{fig:thermal}f). Accretion heating has ceased in these histories, so the excess emission is sustained by heat stored during accretion.

The atmosphere dominates the predicted first-exposure emission, so the deeper history's early excess is mainly thermal. In the band around 2--3~keV, the deeper and shallower histories predict about five and two times the observed counts, respectively (Figure~\ref{fig:spectra}a). Across the five exposures in 2009 and 2015, the deeper prediction is about half of the observed count total, whereas the shallower prediction is close to it (Figure~\ref{fig:spectra}c). The deeper history's early excess and late deficit persist when we omit the bin near 1.4~keV and refit the thermal and spectral parameters (Appendix~\ref{app:spectra}).

The larger fitted absorbing column partly attenuates this early excess but suppresses the later spectra. Replacing only this column with the shallower fitted value, while holding the deeper history's temperature evolution and all other parameters fixed, roughly doubles both the early and late predicted counts. The late deficit decreases as the early excess grows.

We test whether the preference for shallower heating persists at a fixed absorbing column. We adopt the value inferred by \citet{merritt2016} and refit the heating strength, core temperature and first-exposure power law. The cooling-only fits change little. With carbon instability imposed, both histories predict a larger early excess, especially for deeper heating. The deeper history is refitted with a lower core temperature, which reduces its late thermal emission and enlarges the late deficit. It still predicts more than four times the observed first-exposure count total. Fixing the absorbing column therefore worsens both fits without reversing the preference for shallower heating (Appendix~\ref{app:search}).

\begin{table*}[t]
\centering
\caption{Heating fits with two carbon fusion rates at $X_{\rm C}=0.20$.}
\label{tab:alternativerate}
\setlength{\tabcolsep}{6pt}
\renewcommand{\arraystretch}{1.12}
\begin{tabular}{lrrrrrr}
\toprule
 & \multicolumn{2}{c}{CF88} & \multicolumn{2}{c}{T18} & \multicolumn{2}{c}{T18}\\
$y_*$ ($10^{12}$ g\,cm$^{-2}$) & \multicolumn{2}{c}{0.630} & \multicolumn{2}{c}{0.630} & \multicolumn{2}{c}{1.000}\\
\cmidrule(lr){2-3}\cmidrule(lr){4-5}\cmidrule(lr){6-7}
Quantity & Shallower & Deeper & Shallower & Deeper & Shallower & Deeper\\
\midrule
$T_{\rm crit}$ ($10^8$ K) & \multicolumn{2}{c}{6.20} & \multicolumn{2}{c}{5.48} & \multicolumn{2}{c}{5.15}\\
$q_a$ (MeV per nucleon) & 2.19 & 4.00 & 1.78 & 3.25 & 1.59 & 2.86\\
$T_{\rm c}$ ($10^7$ K) & 7.77 & 8.32 & 7.78 & 8.16 & 7.73 & 8.02\\
$N_{\rm H}$ ($10^{22}$ cm$^{-2}$) & 1.59 & 2.48 & 1.46 & 2.12 & 1.40 & 1.94\\
$\Delta W$ & 39 & 352 & 10 & 211 & 3 & 145\\
\bottomrule
\end{tabular}
\end{table*}

\begin{table*}[t]
\centering
\caption{Heating fits at three fuel depths using CF88, with $X_{\rm C}=0.20$.}
\label{tab:eventdepth}
\setlength{\tabcolsep}{5pt}
\renewcommand{\arraystretch}{1.15}
\begin{tabular}{llccc}
\toprule
Quantity & Heating Region & \multicolumn{3}{c}{\shortstack{Fuel Column $y_*$\\($10^{12}$ g\,cm$^{-2}$)}}\\
\cmidrule(lr){3-5}
 & & 0.630 & 0.794 & 1.000\\
\midrule
Instability temperature, $T_{\rm crit}$ ($10^8$ K) & Both & 6.20 & 6.03 & 5.86\\
\addlinespace[4pt]
Heating, $q_a$ (MeV/nucleon) & Shallower & 2.19 & 2.08 & 1.97\\
 & Deeper & 4.00 & 3.76 & 3.54\\
\addlinespace[4pt]
Increase in spectral statistic, $\Delta W$ & Shallower & 39 & 29 & 21\\
 & Deeper & 352 & 306 & 263\\
\bottomrule
\end{tabular}
\end{table*}

\subsection{The Heating Contrast Persists at Lower Instability Temperatures}
\label{sec:carbonrate}
\label{sec:eventdepth}

A lower instability temperature requires less accretion heating and therefore produces a smaller early quiescent excess. At the smallest fuel depth, the faster T18 fusion rate reduces the refitted heat per accreted nucleon by about 20\% relative to CF88 in both regions (Table~\ref{tab:alternativerate}). Deeper heating continues to overpredict the first exposure and underpredict the late spectra, and shallower heating remains closer to the observations (Figure~\ref{fig:spectra}b,d).

With CF88 and $X_{\rm C}=0.20$, increasing the fuel column lowers the instability temperature and the required accretion heating (Table~\ref{tab:eventdepth}). The one-zone conductive loss contains an explicit $y^{-2}$ factor (Equation~\ref{eq:conductiveloss}); a larger overlying column reduces this loss at fixed conductivity, density and temperature. The deeper source still requires substantially more heat per accreted nucleon, and shallower deposition remains closer to the cooling observations at all three depths.

Combining T18 with the largest fuel column lowers the instability temperature further. Shallower heating then fits the spectra nearly as well as it does without the carbon condition. Its predicted count totals exceed the observations by about 10\% in both the first exposure and the final five exposures. In the conditional count simulations, the fraction of realizations with $W$ at least as large as observed is comparable to that for the cooling-only fits (Appendix~\ref{app:spectra}). The deeper history retains a larger early excess, a late deficit and a much higher spectral statistic (Table~\ref{tab:alternativerate}).

\subsection{The Superburst Decay Constrains the Heated Layer}
\label{sec:burstdecay}

With the common carbon abundance inferred from decay fits with uniform energy release at each fuel depth, the refitted CF88 histories give decay fluxes that differ by less than 0.1\% between heating regions (Appendix~\ref{app:uniformrelease}). Because the fuel temperatures are similar, the same prescribed energy release produces nearly identical decays, whereas the quiescent spectra reflect different amounts of heat stored farther inward. Of the three fuel depths, the middle depth gives the best fit to the observed decay.

At this middle depth, the shallower T18 history satisfies local carbon instability and fits the observed decay with uniform energy release and the same carbon abundance in the instability and decay calculations. Its quiescent fit is nearly as good as the cooling-only fit. The deeper history gives essentially the same decay and retains a larger early quiescent excess and late spectral deficit (Appendix~\ref{app:uniformrelease}).

\begin{figure}[t]
\centering
\figfile[width=75mm]{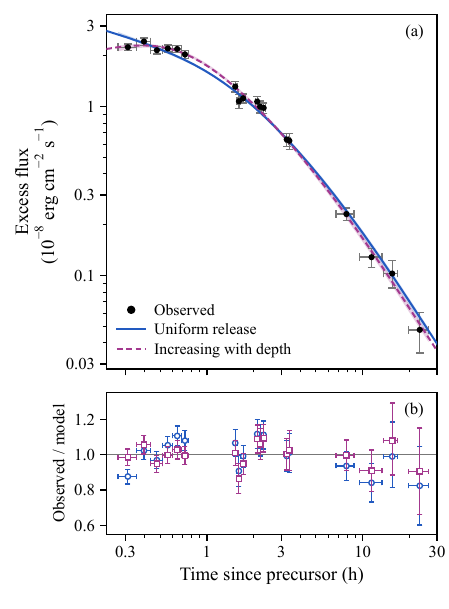}
\caption{Superburst decay on the thermal background from the shallower heating region, refitted at $y_*=7.94\times10^{11}$~g\,cm$^{-2}$ and $X_{\rm C}=0.226$. Points show the observed fluxes from \citet{kuulkers2002}; gray bars show the digitized flux errors and observing intervals. (a) Curves show the selected models; shading shows pointwise conditional 68\% ranges of the model flux (Appendix~\ref{app:releaseprofile}). (b) Observed flux divided by the exposure-averaged model flux for uniform release (open circles) and release increasing with depth (open squares); vertical bars show the digitized flux errors divided by the same model averages.}
\label{fig:burstdecay}
\end{figure}

For the CF88 thermal backgrounds at the middle fuel depth, energy release increasing with depth gives a flatter early decay that follows the measured flux more closely than uniform specific energy release (Figure~\ref{fig:burstdecay}a,b). For both backgrounds, concentrating more energy toward the bottom of the heated layer gives a shallower fitted layer, while the total deposited energy changes by only a few percent (Appendix~\ref{app:releaseprofile}). Fixing the onset at the short X-ray burst that preceded the superburst (i.e., the precursor; \citealt{kuulkers2002}) preserves the shift to a shallower heated layer at nearly unchanged total energy.

At 7~kpc, fitting the decay on the corresponding fixed thermal backgrounds requires more burst energy and a larger burst-heated column. Release increasing with depth reproduces the early flux more closely; both distributions underpredict the emission several hours later by about one-third. The two accretion-heating backgrounds again give nearly identical decays (Appendix~\ref{app:releaseprofile}).

At 4.5~kpc, we test whether models including nuclear Urca cooling can reproduce the observed decay with burst-heated layers deeper than those favored by the preceding fits. Thermal neutrino emission \citep{1996ApJS..102..411I} can remove a substantial fraction of the energy in hot, deep superbursts \citep{2023AstL...49..824K}. Nuclear Urca cycles provide an additional neutrino loss mechanism \citep{2026arXiv260219018H}. The proton branch of carbon fusion produces $^{23}$Na, linking the $^{23}$Na--$^{23}$Ne pair to the burst ashes \citep{tumino2018}. In this pair, electron capture converts sodium to neon, and beta decay returns it to sodium; the emitted neutrinos and antineutrinos carry thermal energy out of the star. The cycle is concentrated near the depth at which electron energies reach the capture threshold, where both reactions can proceed \citep{deibel2016}. The pre-burst thermal histories and their fuel depth $y_*$ remain fixed. We extend burst heating to the capture shell at $y\simeq2.6\times10^{12}$~g\,cm$^{-2}$ and below it, including Urca cooling in the temperature evolution. With the mean deposited energy and onset refitted, the uniform-release models with Urca cooling still predict about three times the observed flux in the final four intervals, on average (Appendix~\ref{app:urca}).

\subsection{Evolving Carbon Allows Weaker Accretion Heating}
\label{sec:evolvingcarbon}

Carbon consumption during burial leaves the adopted fuel layer locally stable at the observed superburst epoch for both fusion rates when the temperature histories are held fixed. Heat released by carbon burning raises the fuel temperature and accelerates fusion. At the original shallower CF88 heating strength, a localized temperature maximum develops below the prescribed fuel layer, and rapid burning begins roughly eight months before the observed superburst (Appendix~\ref{app:coupledburning}).

Including nuclear heating reduces the accretion heating needed for rapid burning to begin near the observed superburst epoch in all four histories. The weaker accretion heating lowers the predicted early thermal excess and the fitted absorbing column, bringing the shallower histories' spectral fits closer to the cooling-only fits. The shallower T18 history nearly matches the first-exposure count total and gives an overall spectral fit comparable to the cooling-only fit. With deeper deposition, the quiescent fits also improve for both rates, but the predicted spectra retain an early excess and a late deficit.

The histories with evolving carbon release nuclear energy comparable to that inferred from the prescribed-energy decay fits, but their burst light curves are broader than observed for both fusion rates. The early flux is too low and the late tail too bright. Rapid burning begins below the fitted burst-heated layer, and much of the released heat remains in the interior after the first few days (Appendix~\ref{app:coupledburning}).

\begin{table*}[t]
\centering
\caption{Heating parameters, required carbon depths and cooling differences for selected synthetic histories.}
\label{tab:synthetic}
\setlength{\tabcolsep}{5pt}
\renewcommand{\arraystretch}{1.15}
\begin{tabular}{lccc}
\toprule
Quantity & Reference & \multicolumn{2}{c}{Deeper Heating}\\
\cmidrule(lr){3-4}
 & & Carbon Beyond $Y_0$ & Carbon Within $Y_0$\\
\midrule
Deposition interval, $\log_{10}P$ & 26--27 & 27--28 & 27--28\\
Heating, $q_a$ (MeV/nucleon) & 2.81 & 2.59 & 4.56\\
Core temperature, $T_{\rm c}$ ($10^7$ K) & 8.00 & 7.89 & 7.24\\
\addlinespace[4pt]
Shallowest unstable depth ($Y_0$) & 0.61 & 4.82 & 0.88\\
Maximum cooling difference (eV) & 0.0 & 1.0 & 21.3\\
\bottomrule
\end{tabular}
\end{table*}

\let\savedsubsection\subsection
\patchcmd{\subsection}{9pt plus 1pt minus 1pt}{9pt}{}{\PackageError{aastex-layout}{Subsection spacing patch failed}{}}
\patchcmd{\subsection}{\center}{\centering}{}{\PackageError{aastex-layout}{Subsection alignment patch failed}{}}
\subsection{Carbon in Older Material Permits a Deeper Heating Alternative}
\let\subsection\savedsubsection
\label{sec:syntheticresults}

In the synthetic comparison, the shallower-heating reference reaches local carbon instability within the newly accreted layer at the selected epoch. A history with deeper heating reproduces the reference effective temperatures to within about 1~eV, but its shallowest locally unstable layer lies at about five times the newly accreted column (Table~\ref{tab:synthetic}). Requiring instability within $Y_0$ for the same deeper heating region instead gives stronger heating and a maximum cooling difference of about 21~eV. The close cooling match with deeper heating therefore depends on carbon in material that predates the accretion episode.

\section{Interpreting the heating location}
\label{sec:discussion}

With prescribed fuel, the mismatch that \citet{meisel2024} identified between carbon ignition and quiescent cooling depends on where heat is deposited as well as on the fusion rate (Sections~\ref{sec:ksfits}--\ref{sec:burstdecay}).

\subsection{Carbon Accumulation and Survival}

The integrated accretion history determines whether fuel at a given depth was accreted during the current outburst or earlier. At 4.5~kpc, the depths tested at and below the Urca shell exceed the newly accreted column, so part of the heated layer lies in older material. At 7~kpc, the same depths fall within the newly accreted column (Appendix~\ref{app:history}). For the synthetic deeper-heating alternative to apply to KS, carbon must survive at the depth where instability develops. The burned fraction and energy distribution must then reproduce the observed decay on the thermal background of a history that fits the quiescent spectra.

Whether another superburst can occur before accretion ends depends on carbon production, survival during burial and replenishment after burning \citep{cooper2006,keek2016}. In the four histories with evolving carbon, only about $10^{12}$~g\,cm$^{-2}$ is accreted after the first burning episode, leaving most newly supplied carbon above the layers that first burned at $1.5$--$1.7\times10^{12}$~g\,cm$^{-2}$. The declining accretion rate cools this replenished fuel, and none of the histories develops a second rapid temperature rise before accretion ends. The superburst recurrence interval of KS has not been measured \citep{kuulkers2002}.

\subsection{Carbon Burning and the Burst Decay}

In the initial comparison with prescribed fuel, we equate the local instability depth with the bottom of the burst-heated layer (Section~\ref{sec:heatingcriteria}), linking ignition at one depth to the extent of energy deposition. Burning can spread beyond its starting point, and heat transport then shapes the decay. The burst light curve can constrain how carbon burns: the radial temperature profile inferred for 4U~1636$-$536 is inconsistent with instantaneous local burning \citep{keek2015}. The multizone models of \citet{2012ApJ...752..150K} and \citet{nava2026superbursts} start from a prescribed carbon-rich supply and follow nuclear reactions through ignition and the burst to calculate the energy release.

In our evolving-carbon calculations, rapid burning begins below the layer favored by the decay fits (Section~\ref{sec:evolvingcarbon}). Further inward burning in the CF88 histories deposits energy still deeper, while the observed decay favors shallower energy deposition. These calculations include only conductive heat transport (Section~\ref{sec:decaymethod}). They differ from the prescribed-energy calculations in the time dependence of nuclear heating, initial thermal state and envelope treatment.

\subsection{Heat Sources and Transport}

The heating needed to warm the fuel and reproduce quiescent emission depends on crustal reactions and energy losses. Neutron transfer changes reaction chains, modifying both heat release and neutrino losses \citep{2019MNRAS.483L..47C,2022ApJ...925..205S}, while Urca cycles in the crust carry energy away in neutrinos \citep{2014Natur.505...62S}. Heat supplied by H/He burning through the outer boundary can reduce the additional shallow heating inferred from the cooling of KS and other transient sources \citep{nava2026cooling}. This contribution could account for part of the heat attributed to the adjustable crustal source in the static-envelope KS fits.

\citet{nava2026superbursts} found that changes in base heating can produce effects on burst timescales comparable to those of changes in the fusion rate. Their superburst models supply heat through a prescribed luminosity at the base of the envelope. Our adjustable source instead deposits heat within the crust, so its depth can be tested against both the ignition condition and the subsequent quiescent emission.

Interpreting quiescent emission in terms of stored heat also depends on how that heat is transported through the crust. Thermodynamically consistent crust models that allow free neutrons to redistribute change the shallow heating and impurity parameters inferred from the cooling of MXB~1659$-$29 \citep{potekhin2023}.

\subsection{Application to Other Transient Sources}

Other transient sources have shown superbursts and quiescent cooling, along with evidence for fuel inherited from earlier outbursts. Aql~X-1 produced a superburst during its 2013 outburst, followed by quiescent thermal emission consistent with crust cooling \citep{alizai2023,waterhouse2016}; its later cooling also depends on heat retained between accretion episodes \citep{ootes2018}. In 4U~1608$-$522, the 2005 superburst occurred 55 days into an outburst, yet its decay implied a fuel column larger than could have accumulated during that interval \citep{keek2008}. Applying this comparison to these systems would test which heating depths and strengths yield a thermal history that brings surviving carbon to instability and reproduces the subsequent quiescent spectra.

\vspace{2.5pt}
\section{Conclusions}
\label{sec:conclusions}
\setlength{\baselineskip}{12.5pt}\setlength{\parskip}{0pt}

{\ifdefined\pdffontexpand\pdffontexpand\font 20 20 1 autoexpand\pdfadjustspacing=2\fi
\looseness=1 We used the superburst and quiescent cooling of KS~1731$-$260 to compare heating that overlaps the carbon fuel layer with heating confined below it. Without the carbon instability condition, histories with either heating region fit the quiescent spectra similarly. Requiring local carbon instability favors heating that overlaps the fuel. This preference persists across the tested fuel depths, fusion rates, distances, absorption treatments and carbon evolution. Heating the fuel from below requires a stronger source and leaves more stored heat to sustain excess early quiescent emission. For the same prescribed burst energy release, the two histories have similar decays despite this difference in stored heat.\par}

The inferred heating location also depends on where carbon survives and how it burns. In the synthetic comparison, carbon in older material permits a deeper-heating alternative with nearly the same cooling as the shallower-heating reference. In the KS histories, including carbon burial, consumption and nuclear heating reduces the accretion heating needed for rapid burning near the observed superburst, but the conductive calculations still produce broader burst emission than observed. Heating models for accreting neutron stars should therefore account for ignition, burst emission and quiescent cooling with a consistent fuel distribution and thermal history.

\makeatletter
\@booleanfalse\raggedcolumn@sw
\onecolumngrid
\makeatother
\setlength{\baselineskip}{12.5pt plus 0.2pt}
\begin{acknowledgments}
This work is supported by JSPS KAKENHI (Grant No. JP25K17403).
This research used observations from the \href{https://cxc.harvard.edu/cda/}{Chandra Data Archive} and source positions from the \href{https://cxc.harvard.edu/csc/}{Chandra Source Catalog}, both provided by the Chandra X-ray Center (CXC). We also used data and software from the \href{https://heasarc.gsfc.nasa.gov/}{High Energy Astrophysics Science Archive Research Center} (HEASARC) at NASA/GSFC. The Chandra data collection is available at \dataset[doi:10.25574/cdc.727]{https://doi.org/10.25574/cdc.727}.

ChatGPT 6 (OpenAI; \url{https://chatgpt.com/}) was used to improve the language and logical flow of the manuscript and to assist with the development and execution of some simulation and \mbox{analysis code}.
\end{acknowledgments}

\appendix
\setlength{\parskip}{0pt}
\raggedbottom
\setlength{\textfloatsep}{9pt minus 1pt}
\renewcommand{\theHequation}{\thesection.\arabic{equation}}
\section{Thermal evolution and spectral fitting}
\label{app:implementation}

\subsection{Thermal Structure and Heating}

In the prescribed HZ90 crust used for the KS thermal evolution, the layer at $y_*$ is iron, while deeper layers contain the evolved nuclei and eventually free neutrons \citep{haensel1990}. We set the dimensionless impurity parameter, the ion-fraction-weighted variance of nuclear charge, to 4, following the low-impurity case of \citet{meisel2024}. Each history begins with a constant redshifted temperature set by its local core temperature. The accretion and quiescent histories omit shell Urca emission; Appendix~\ref{app:urca} treats this loss during burst cooling.

For KS, the standard outer and inner crustal sources have amplitudes of 0.3 and 1.5~MeV over $\log_{10}P=26.86$--30.13 and 30.42--31.2. The fitted source heats entire radial zones whose central pressures lie within its deposition interval. For heating region $a$, let $M_a$ be the total mass of these zones, $Q_a$ the source normalization in MeV, and $m_u$ the atomic mass unit. With $Q_a$ converted to erg, the local heating rate per unit mass $\epsilon_a$ is
\begin{equation}
 \epsilon_a=\frac{\mdot Q_a}{M_a m_u}.
\end{equation}
Local proper time $\tau$ and observer time $t$ are related by $d\tau=e^\Phi dt$, where $\Phi$ is the dimensionless metric potential. With zone mass $\Delta m_i$ and metric potential $\Phi_i$, the local heat deposited per accreted nucleon is
\begin{equation}
q_a=Q_a\frac{\sum_{i\in a} e^{\Phi_i}\Delta m_i}{\sum_{i\in a}\Delta m_i}.
\end{equation}
The envelope light-element column is $10^4$~g\,cm$^{-2}$ during both accretion and quiescence \citep{brown2009}.

For the KS heating fits, the fixed grid has 345 zones over $\log_{10}P=25$--32.5, with target spacing $\Delta\ln P=0.05$; each heating interval extends to the faces enclosing the selected zones. Thermal evolution is integrated with adaptive time steps and tolerance $10^{-4}$. Temperatures at a specified depth are interpolated in $\log T$ and $\log P$, while the conductive luminosity $L^\infty$ is the local luminosity at each zone boundary multiplied by $e^{2\Phi}$. Heat transport follows the gradient of $e^\Phi T$, so a maximum of local $T$ need not coincide exactly with zero flux.

For the two CF88 histories fitted with carbon instability in Table~\ref{tab:ksfits}, we retain the original heating boundaries and heat per accreted nucleon. We tighten the integration tolerance to $10^{-6}$ and then increase the grid from 345 to 690 zones. With the thermal and spectral parameters fixed, the fuel temperature at the superburst changes by less than 0.07\% and the predicted count total in each exposure, including scaled measured background, by less than 0.9\%. We then increase $q_a$ by 0.10\% for the shallower history to satisfy Equation~\ref{eq:ignition} on the finer grid; the deeper history satisfies the condition at its original $q_a$. Refitting only the absorbing column and first-exposure power-law amplitude gives a deeper-minus-shallower difference in $W$ of 312.4, compared with 311.9 on the original grid.

In the synthetic comparison, the fitted and standard heat sources add where their regions overlap. The local heating rate includes a factor $e^{-\Phi}$ to convert the observer-frame accretion rate to mass per unit proper time, so $q_a=Q_a$. During accretion, we evaluate the tabulated burning envelope of \citet{nava2025} at a base density of $10^7$~g\,cm$^{-3}$ using the local flux and accretion rate; after accretion ceases, we use the static envelope.

For the synthetic reference, we adopt $M=1.4\,M_\odot$, $R=11.56$~km and $T_{\rm c}=8.0\times10^7$~K. The reference has a constant accretion rate of $10^{17}$~g\,s$^{-1}$ and heating at $\log_{10}P=26$--27 with $q_a=2.807$~MeV per accreted nucleon. The carbon instability condition is evaluated 2983.92~d after accretion begins, when the newly accreted column is $Y_0=1.54\times10^{12}$~g\,cm$^{-2}$. The corresponding accumulation time is $4\pi R^2Y_0/\mdot\simeq8.2$~yr in the observer frame. Accretion ends at 4565.625~d, when we set the rate to zero and sample cooling 64.64, 100, 300, 750, 1500, 3000 and 5300~d later. The composition remains prescribed throughout.

\subsection{Local Carbon Microphysics}
\label{app:material}

At each $(P,T)$ in a thermal history, we recalculate the carbon--iron fuel density with the thermal model's equation of state. The screened CF88 rate includes a factor of one half for identical reactants, with 13.9336~MeV released per carbon pair for conversion into $^{24}$Mg.

Let $K$ be the thermal conductivity and $\rho$ the mass density, expressed in units of erg\,cm$^{-1}$\,s$^{-1}$\,K$^{-1}$ and g\,cm$^{-3}$, respectively. The one-zone estimate of the conductive cooling rate is
\begin{equation}
 \epsilon_{\rm cool}=0.17\,\frac{K\rho T}{y^2}.
 \label{eq:conductiveloss}
\end{equation}
The factor 0.17 lies within the range calibrated by \citet{potekhin2012} for their envelope models. 

For the conductivity calculation, let $n_e$ denote electron number density, $\mu_e$ the electron Fermi energy including rest energy, and $\nu_{ei}$ the electron--ion collision frequency. For ionic species indexed by $i$, the mass fraction, mass number, charge number and Coulomb logarithm are denoted by $X_i$, $A_i$, $Z_i$ and $\Lambda_i$, respectively. We use $c$ for the speed of light, $k_B$ for the Boltzmann constant, $\hbar$ for the reduced Planck constant, $m_e$ for the electron mass and $\alpha$ for the fine-structure constant. The liquid thermal conductivity is represented by
\begin{equation}
 K=\frac{\pi^2c^2k_B^2Tn_e}{3\mu_e\nu_{ei}},\qquad
 \nu_{ei}=\frac{4\alpha^2\mu_e}{3\pi\hbar}
 \frac{\sum_i (X_i/A_i)Z_i^2\Lambda_i}{\sum_i (X_i/A_i)Z_i}.
\end{equation}
The electron number density is $n_e=\rho\sum_i X_iZ_i/(A_i m_u)$, and the corresponding Fermi energy is $\mu_e=[m_e^2c^4+\hbar^2c^2(3\pi^2n_e)^{2/3}]^{1/2}$. We adopt $\Lambda_i=1$. This liquid electron--ion approximation has the form used by \citet{meisel2024}, with the ionic number weighting of the mixture scattering sum \citep{cassisi2007} and no separate lattice impurity resistance.

Temperature derivatives are calculated by centered logarithmic differences with step $10^{-3}$, reevaluating density, conductivity and screening at fixed pressure. For the histories fitted with local carbon instability, Equation~\ref{eq:ignition} remains satisfied with this step and after halving it. The synthetic comparison uses the same conductivity prescription with the corresponding gravity and pressure.

For the T18 fits in Table~\ref{tab:alternativerate}, we replace CF88 with the unscreened total fusion rate tabulated by \citet{tumino2018}, retaining the fuel mixture, equation of state, screening, conductivity and energy per reaction. At $T=0.5$~GK, T18 is about 25 times CF88, reflecting the low-energy resonances inferred from their Trojan Horse measurement. We use shape-preserving cubic interpolation in logarithmic temperature and rate and recalculate the temperature derivatives at fixed pressure. Cubic spline and linear interpolation in the same coordinates change the instability temperature at $y_*=6.3\times10^{11}$~g\,cm$^{-2}$ by less than 0.4\%.

\subsection{Carbon Burial and Nuclear Heating}
\label{app:coupledburning}

Carbon supplied at a smaller column can be consumed before reaching the adopted fuel depth $y_*$. Let $r_{\rm CC}$ be the number of carbon fusion reactions per gram per proper second.

Along the trajectory of an accreted fluid element, advection and carbon consumption satisfy
\begin{equation}
 \frac{dy}{dt}=\frac{\mdot(t)}{4\pi R^2},\qquad
 \frac{dX_{\rm C}}{dt}=-24m_u r_{\rm CC}e^\Phi.
\end{equation}
We represent the fusion ashes as $^{24}$Mg, with the iron fraction unchanged.

We first follow carbon consumption in material reaching $y_*=6.3\times10^{11}$~g\,cm$^{-2}$ at the observed superburst epoch, $t_{\rm sb}$, in both histories selected with $X_{\rm C}=0.20$ in Table~\ref{tab:ksfits}. Material is supplied at $y=10^{11}$~g\,cm$^{-2}$ with carbon and iron mass fractions $X_{\rm C}=0.20$ and $X_{\rm Fe}=0.80$, respectively. Its entry time is determined by the prescribed accretion history. Density and the screened rate follow the evolving mixture, while temperature and heat flow are prescribed by the fitted histories without feedback from carbon burning. With the integration step limited to two days, the carbon fraction falls to 0.17--0.18 after about 700~d of burial; halving this limit leaves the final carbon fraction unchanged at the quoted precision in both histories.

For T18, the passive burial calculation uses the histories refitted at the middle fuel depth with the common abundance $X_{\rm C}=0.227$ (Appendix~\ref{app:uniformrelease}). Material supplied at this abundance reaches the adopted fuel depth after about 940~d with $X_{\rm C}=0.208$ and 0.211 in the shallower and deeper histories, respectively. Halving the integration step limit from two days to one changes the final carbon fractions by less than $5\times10^{-8}$. At the observed superburst epoch, the depleted mixtures are locally stable at the adopted fuel depths on all four prescribed temperature histories.

To include heat from carbon burning, we couple carbon advection and consumption to the dStar temperature evolution. We use the shallower CF88 history fitted with local carbon instability at the smallest adopted fuel depth (Table~\ref{tab:ksfits}) and retain its accretion history, heating strength, core temperature and thermal boundaries. The calculation begins at the start of the adopted accretion episode with the fitted history's isothermal redshifted temperature and no carbon in the crust. Material entering at $y=1.03\times10^{11}$~g\,cm$^{-2}$ contains 20\% carbon and 80\% iron, representing carbon production above this depth. Below it, carbon and its magnesium product are advected conservatively without a prescribed lower fuel boundary.

Carbon consumption and nuclear heat release are calculated using the same CF88 reaction rate, with Coulomb screening evaluated for the evolving mixture. The adopted conversion to $^{24}$Mg releases $5.60\times10^{17}$~erg per gram of carbon consumed, excluding the energy from subsequent reactions to iron-group nuclei. The larger energy yield used for complete burning in the separate decay model (Appendix~\ref{app:burstdecay}) includes those later stages. Hydrostatic density and the transport composition remain those of the prescribed HZ90 crust. The heat capacity, conductivity and neutrino emissivity retain their temperature dependence.

\begin{figure*}[t]
\centering
\figfile{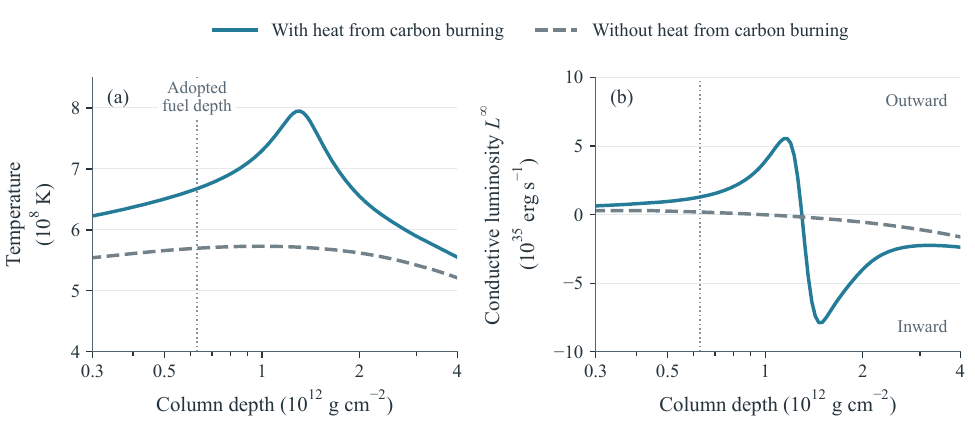}
\caption{Temperature (a) and conductive luminosity (b) during carbon burial, with and without heat from carbon burning, for the fitted shallower accretion source. Profiles are shown at the same time during the rapid temperature rise, when the calculation including carbon heating has a peak temperature of approximately $8\times10^8$~K. Both calculations have the same initial state and carbon supply. Vertical dotted lines mark the adopted fuel depth used in the heating fits. }
\label{fig:carbonheating}
\end{figure*}

Continuously supplied carbon produces a localized temperature maximum near $y=1.3\times10^{12}$~g\,cm$^{-2}$, about twice the adopted fuel depth. Heat from carbon burning raises the temperature above the broad accretion-heated profile (Figure~\ref{fig:carbonheating}a), with heat flowing outward above the peak and inward below it (Figure~\ref{fig:carbonheating}b).

When the heat released by carbon burning is omitted, the temperature at $t_{\rm sb}$ differs from the original fitted background by less than 0.07\%. For the calculation including carbon heating, we increase the grid from 345 to 690 zones and tighten the thermal and abundance tolerances to $10^{-6}$ and $10^{-7}$. The resulting peak depths agree within 1.3\% at a peak temperature of $8\times10^8$~K. Both grids produce a temperature excess of about $2.3\times10^8$~K and outward and inward heat flow on opposite sides of the peak. The supply boundary and total accretion heating remain fixed; the discretized heating boundaries shift by about 2.5\%. Figure~\ref{fig:carbonheating} uses the finer grid.

The rapid temperature rise follows the transition from the assumed constant premonitoring accretion rate to the ASM reconstruction. At this transition, the adopted rate increases by about 24\%, and carbon supplied during the preceding evolution is already present below $y_*$. The stronger accretion heating is accompanied by a rising nuclear heating rate, which amplifies the temperature increase; the peak reaches $8\times10^8$~K about 16~d later. With the heating parameters fixed at the values fitted with prescribed fuel, the timing of the rise depends on the adopted premonitoring history and fuel supply.

After the snapshot in Figure~\ref{fig:carbonheating}, the peak temperature continues to rise, crossing $10^9$~K in about 100~s of observer time. The calculation stops at this prescribed temperature limit, while the fuel is still heating rapidly.

The weaker-heating CF88 histories in Table~\ref{tab:evolvingcarbon} use the 690-zone grid and continuous carbon supply. For CF88, we bracket the heating strength required for the first crossing of $10^9$~K to occur by $t_{\rm sb}$. We narrow the bracket to less than 2\% of the prescribed-fuel strength and retain the history that crosses before the event. Nuclear evolution follows the conversion of carbon into magnesium, without subsequent reactions, an imposed energy pulse or a carbon reset. The relative temperature and abundance tolerances are $10^{-6}$ and $10^{-7}$, respectively, with absolute abundance tolerance $10^{-10}$. To follow the brief increase in envelope base temperature after burning, we use the original equations to extend the static helium--iron envelope relation to higher temperatures. The tabulated values and derivatives remain unchanged over the original range.

We also evolve carbon and nuclear heating with T18, starting from the two thermal configurations fitted at the middle fuel depth with prescribed $X_{\rm C}=0.227$ (Appendix~\ref{app:uniformrelease}). Each calculation supplies material with $X_{\rm C}=0.227$ to an initially carbon-free crust. The structure, grid, transport and accretion history are those of the coupled CF88 calculations. Carbon consumption and nuclear heating follow the T18 rate within its tabulated range, $0.1$--$3$~GK. Outside this range, we use CF88 normalized to the T18 rate at the nearest endpoint, preserving continuity.

In Table~\ref{tab:evolvingcarbon}, $t_9$ is the first time a carbon-bearing zone reaches $10^9$~K, and $y_9$ is its column depth. The time difference $t_9-t_{\rm sb}$ is expressed in the observer frame. For T18, we vary only the accretion heating strength between zero and its prescribed-fuel value. We evaluate at most seven histories per region and select a history with $t_9$ within 30 observer-frame days before or after $t_{\rm sb}$. Relative to the prescribed-fuel fits, the selected accretion heating is 17\% weaker for shallower deposition and 20\% weaker for deeper deposition. The early and late predicted-to-observed count ratios use the first exposure and the sum of the final five exposures, respectively, including the same scaled measured background as Figure~\ref{fig:timeline}c.

For each weaker-heating history, we calculate the nuclear energy from the carbon mass consumed, accounting for advective transport between radial cells. During the first three observer-frame days after $t_9$, the local release is approximately $2.8\times10^{42}$~erg for CF88 and $3.2\times10^{42}$~erg for T18, compared with about $2.2\times10^{42}$~erg deposited in the prescribed-energy decay fits. The energy calculated from consumption agrees with the time integral of the nuclear heating rate to within 1\%. In the CF88 histories, about one-quarter of the released energy comes from below $y_9$, showing that burning extends deeper than the layer that first heats rapidly. In the T18 histories, about half of the nuclear energy is released at temperatures above the tabulated upper limit of 3~GK, where the rate follows the normalized CF88 extension.

To determine how the nuclear energy is divided between stored heat, surface radiation and neutrino emission, we compare each weaker-heating history with a second evolution. This evolution starts from the same temperature and composition just before the rapid rise and omits subsequent nuclear heating. It retains the carbon supply, advection and consumption equations, accretion heating and transport prescriptions. We integrate the thermal equation from the common pre-rise state to three observer-frame days after $t_9$. For either fusion rate, approximately two-thirds of the additional nuclear energy remains stored as thermal energy, about one-third emerges as excess surface radiation, and a few percent is lost to additional neutrino emission. These energy differences are evaluated in the distant-observer frame.\par We subtract the redshifted surface luminosity of the evolution without nuclear heating from that of the history with nuclear heating. We convert the difference to flux and average over the 18 observed exposures. For the decay comparison, we shift each calculated light curve in observer time, allowing $t_9$ to fall between the precursor and the first measured interval. All four fits place $t_9$ at the precursor. Table~\ref{tab:evolvingcarbon} reports the original calendar-time offsets of $t_9$ from the observed superburst. The mean predicted-to-observed flux ratios over the first six and last four exposures are about 0.11 and 1.6 for CF88, and 0.15 and 1.85 for T18. The fluence summed over the measured exposures is about 85\% of the observed value for CF88 and agrees within 1\% for T18.

\begin{table*}[t]
\centering
\caption{Selected weaker-heating histories with evolving carbon at 4.5~kpc.}
\label{tab:evolvingcarbon}
\begin{tabular}{lrrrr}
\toprule
 & \multicolumn{2}{c}{CF88} & \multicolumn{2}{c}{T18}\\
\cmidrule(lr){2-3}\cmidrule(lr){4-5}
Quantity & Shallower & Deeper & Shallower & Deeper\\
\midrule
Supplied $X_{\rm C}$ & 0.20 & 0.20 & 0.227 & 0.227\\
$q_a$ (MeV per nucleon) & 1.76 & 3.05 & 1.37 & 2.41\\
$T_{\rm c}$ ($10^7$ K) & 7.77 & 8.32 & 7.76 & 8.05\\
$t_9-t_{\rm sb}$ (d) & $-32$ & $-9$ & $+2$ & $+7$\\
$y_9$ ($10^{12}$ g\,cm$^{-2}$) & 1.52 & 1.68 & 1.60 & 1.72\\
$N_{\rm H}$ ($10^{22}$ cm$^{-2}$) & 1.46 & 2.09 & 1.36 & 1.79\\
Spectral statistic $W$ & 277 & 451 & 267 & 352\\
Early predicted/observed counts & 1.18 & 1.74 & 1.02 & 1.51\\
Late predicted/observed counts & 1.04 & 0.73 & 1.16 & 0.85\\
\bottomrule
\end{tabular}
\end{table*}

\subsection{Searching the Allowed Thermal Histories}
\label{app:search}

For each KS heating region, we begin with a $5\times5$ grid over $0\leq Q_a\leq10$~MeV and $3\times10^7\leq T_{\rm c}\leq1.5\times10^8$~K. To control computational cost, we prescribe limits on additional thermal evolutions; previously calculated histories are reused and do not count toward these limits. We continue the cooling-only fits with the Nelder--Mead method from two nearby starting points, allowing up to 24 parameter trials per starting point and 40 new thermal histories per heating region. The search for each heating region reaches the 40-history limit. These additional evaluations lower the minimum $W$ by 0.022 for shallower heating and 0.026 for deeper heating.

For fits requiring carbon instability, thermal evolution at each core temperature determines the heating strength at equality in Equation~\ref{eq:ignition}. On the initial grid, the fuel temperature increases with heating strength at fixed core temperature, bracketing one crossing of each instability threshold. We sample this boundary and stronger heating at nearby core temperatures, using interpolation to select trial parameters and calculating $W$ through thermal evolution and spectral fitting for each new history. Each initial CF88 search allows up to 28 new thermal histories, including up to four inside the allowed region. A local search for shallower heating at the middle fuel depth evaluates five additional histories within a limit of eight.

For each heating region, the initial T18 search allows up to 16 new thermal histories along the instability boundary and four inside the allowed region. The deeper search reaches this 20-history limit. The shallower search stops after 11 new histories when refinement returns to the same feasible history with the lowest $W$; four additional histories sample stronger heating.

\begin{figure}[!ht]
\centering
\includegraphics[width=0.94\linewidth]{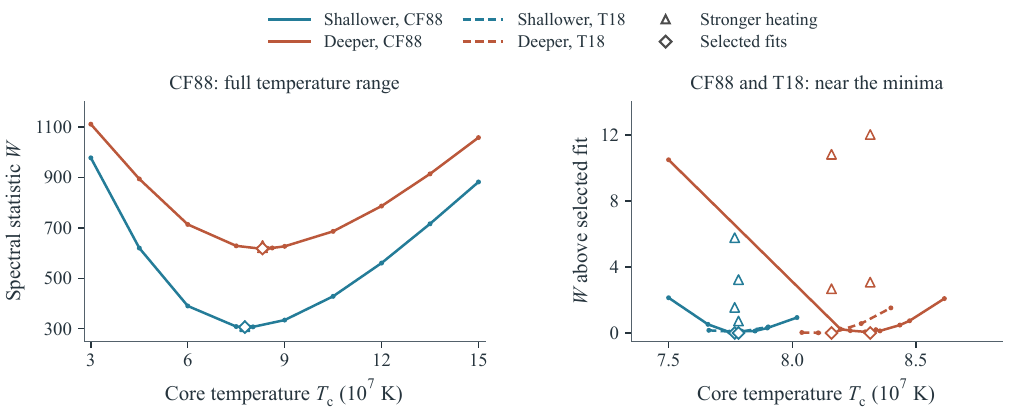}
\caption{Spectral fits along the carbon instability boundary at $y_*=6.3\times10^{11}$~g\,cm$^{-2}$ and $X_{\rm C}=0.20$. Lines join evaluated boundary histories; open triangles denote stronger heating at fixed core temperature, and diamonds mark the selected fits. The left panel shows the full sampled range of core temperatures for CF88; the right shows changes in $W$ relative to each selected fit near the CF88 and T18 minima.}
\label{fig:searchcoverage}
\end{figure}

To test for lower $W$ away from the selected CF88 fits, we extend the boundary sampling at $y_*=6.3\times10^{11}$~g\,cm$^{-2}$ and $X_{\rm C}=0.20$ to nine equally spaced core temperatures across the full search interval, retaining the more closely spaced samples near each minimum. At each additional boundary point, we adjust the heating strength until the fuel temperature lies within 0.001\% above the instability temperature and refit the absorption and first-exposure power-law amplitude. The sampled $W$ rises on both sides of the selected core temperatures (Figure~\ref{fig:searchcoverage}, left).

Using this temperature tolerance, we locate the T18 instability boundary independently of $W$. The selected shallower fit satisfies Equation~\ref{eq:ignition} at a temperature slightly above $T_{\rm crit}$ and has a lower $W$ than the boundary history at the same core temperature. New boundary histories at core temperatures $1.2\times10^6$~K below and above the selected value also have higher $W$ (Figure~\ref{fig:searchcoverage}, right). At $T_{\rm c}=6\times10^7$ and $1.2\times10^8$~K, the additional boundary histories have higher $W$ for both heating regions. Stronger heating at the selected core temperatures also raises $W$.

For CF88, we repeat the constrained search at the smallest fuel depth using the abundance fitted to the burst decay (Appendix~\ref{app:uniformrelease}; Table~\ref{tab:depthdecay}). Evaluated histories at higher and lower core temperatures and inside the allowed region have larger $W$, and both the original and refitted histories satisfy Equation~\ref{eq:ignition} at this abundance. We report the lowest-$W$ fits satisfying Equation~\ref{eq:ignition} among those evaluated in these finite searches.

For T18 at $y_*=10^{12}$~g\,cm$^{-2}$, we sample five boundary histories over $6\times10^7\leq T_{\rm c}\leq1.2\times10^8$~K. We refine near each minimum and evaluate four histories with stronger heating. Each search uses 35 histories within a limit of 40 new thermal evolutions per region. The shallower refinement reaches its limit of eight local trials; histories evaluated at core temperatures on either side of the selected value, as well as those with stronger heating, have higher $W$.

For the 7~kpc comparison, we change the atmosphere normalization and multiply the rates throughout the adopted accretion history by $(7/4.5)^2$, including the assumed rate before ASM monitoring. We refit the thermal and spectral parameters with $X_{\rm C}=0.20$ and CF88. Separate decay fits on the resulting backgrounds vary the burst-heated depth and released energy (Appendix~\ref{app:releaseprofile}). Another comparison retains 4.5~kpc and fixes $N_{\rm H}=1.30\times10^{22}$~cm$^{-2}$ while refitting the remaining parameters. This fixed-absorption benchmark comes from the Chandra and XMM--Newton cooling data of \citet{merritt2016}, which overlap the Chandra observations fitted here.

Both comparisons sample the carbon instability boundary at nine core temperatures across the original range, refine near each minimum and evaluate four histories with stronger heating. The distance and fixed-absorption searches allow at most 150 and 125 new thermal histories per region, respectively. Continuing the 7~kpc cooling fits changes $W$ by less than 0.002. Table~\ref{tab:controls} lists the fitted parameters and spectral statistics.

\begin{table}[!ht]
\centering
\caption{Fits requiring carbon instability with a larger distance or a fixed absorbing column.}
\label{tab:controls}
\begin{tabular}{lrrrr}
\toprule
 & \multicolumn{2}{c}{7 kpc} & \multicolumn{2}{c}{Fixed $N_{\rm H}$}\\
\cmidrule(lr){2-3}\cmidrule(lr){4-5}
Quantity & Shallower & Deeper & Shallower & Deeper\\
\midrule
$q_a$ (MeV per nucleon) & 0.87 & 1.59 & 2.21 & 4.10\\
$T_{\rm c}$ ($10^7$ K) & 8.93 & 9.08 & 7.07 & 5.62\\
$N_{\rm H}$ ($10^{22}$ cm$^{-2}$) & 1.34 & 1.87 & 1.30 & 1.30\\
$W$ & 277 & 453 & 333 & 1068\\
$\Delta W$ & 15 & 191 & 66 & 801\\
\bottomrule
\end{tabular}
\end{table}

The synthetic histories span $0\leq q_a\leq5$~MeV per accreted nucleon and $5\times10^7\leq T_{\rm c}\leq1.5\times10^8$~K. The fit minimizes the equally weighted sum of squared temperature differences, treating the reference temperatures as noiseless data and expressing temperatures in eV. 

For these histories, we search for local carbon instability over $10^{11}\leq y\leq10^{13.5}$~g\,cm$^{-2}$. We interpolate the thermal profile and instability temperature in $\log T$ and $\log y$ at the depths sampled by the thermal evolution and instability-temperature calculations. Histories that already satisfy the carbon instability condition at the minimum column in the search are excluded. Near each selected crossing, we recalculate the carbon heating and conductive cooling rates and their temperature derivatives.

\subsection{Cooling Spectra}
\label{app:spectra}

We reprocess the archival Chandra observations with CIAO~4.18.0 \citep{fruscione2026} and CALDB~4.12.4, and fit the spectra with XSPEC~12.15.1. The observation identifiers are 2428, 3796, 3797, 6279, 5468, 10037, 10911, 16734, 17706 and 17707. The catalogued source position is transformed to each reprocessed image without recentering. We extract source counts within a $3''$ circle at that position and background counts from a concentric annulus with radii $7''$ and $25''$.

For the near-axis source in each observation, we calculate the energy redistribution and effective collecting area at the source position. The effective area is multiplied once by the energy-dependent fraction of the point spread function enclosed by the extraction aperture; background scaling uses the measured exposures and extraction areas. Source and background spectra have matching groups of eight pulse-invariant channels, selected over approximately 0.5--7~keV, with the same selection and grouping for both heating regions and for Figure~\ref{fig:spectra}.

For the absorption model described in Section~\ref{sec:observations}, we use the solar abundances of \citet{anders1989} and the cross sections of \citet{balucinska1992}, including the updated helium cross section of \citet{yan1998}. The first-exposure power law has fixed photon index 2. Source and background counts enter $W$ as independent Poisson measurements, with the background intensity in each channel fitted jointly to both.

The common absorbing column is bounded by $0.1\leq N_{\rm H}/(10^{22}\,{\rm cm^{-2}})\leq5$, and the power-law photon flux density at 1~keV ranges from zero to $10^{-3}$~photons\,cm$^{-2}$\,s$^{-1}$\,keV$^{-1}$. When absorption is free, each spectral fit starts from three initial values of $N_{\rm H}$; the fixed-absorption comparison instead uses three initial power-law amplitudes.

We assess count fluctuations for each selected thermal history with 300 simulated pairs of source and background spectra. The 540 groups contain 390 observed source-region counts and 1061 background-region counts. Independent Poisson draws use the fitted source prediction and the maximum-likelihood background mean in each group, with the measured exposure and area factors. Each realization is fitted again for the common absorbing column and first-exposure power-law amplitude while holding the thermal history fixed. Table~\ref{tab:countsimulation} lists how many realizations have $W$ at least as large as observed. These comparisons are conditional on the selected thermal history and fitted background spectrum; 123 groups have zero fitted background mean and therefore contribute no background fluctuations in this calculation.

\begin{table}[htbp]
\centering
\caption{Count simulations conditional on selected thermal histories.}
\label{tab:countsimulation}
\begin{tabular}{llrrrr}
\toprule
 & & \multicolumn{2}{c}{Shallower} & \multicolumn{2}{c}{Deeper}\\
\cmidrule(lr){3-4}\cmidrule(lr){5-6}
Fuel condition & $y_*$ ($10^{12}$ g\,cm$^{-2}$) & $W$ & Exceedances & $W$ & Exceedances\\
\midrule
Cooling only & --- & 267 & 31/300 & 266 & 37/300\\
CF88, $X_{\rm C}=0.20$ & 0.630 & 306 & 0/300 & 618 & 0/300\\
T18, $X_{\rm C}=0.20$ & 1.000 & 269 & 25/300 & 412 & 0/300\\
\bottomrule
\end{tabular}
\end{table}

To assess how the excess counts near 1.4~keV affect the heating comparison (Figure~\ref{fig:spectra}c,d), we omit the same grouped bin, 1.4016--1.5184~keV, from all ten spectra. We then refit $q_a$, $T_{\rm c}$, $N_{\rm H}$ and the first-exposure power-law amplitude. At $y_*=6.3\times10^{11}$~g\,cm$^{-2}$ and $X_{\rm C}=0.20$, the deeper-minus-shallower difference in $W$ is 297 with CF88 and 188 with T18, compared with 312 and 201 when the bin is retained; each difference uses identical energy selections for the two regions. Across the four constrained fits, $q_a$ changes by less than 0.4\%, and the core temperature and spectral parameters by less than 4\%. The cooling-only fits remain nearly equivalent, with their $W$ values differing by 0.34.

Each annual ratio in Figure~\ref{fig:timeline}c is placed at the exposure-weighted mean of the contributing midpoints, with the full date span shown. The count uncertainties are displayed with the spectra in Figure~\ref{fig:spectra}.

\section{Reconstructing the accretion history and choosing fuel depths}
\label{app:history}

We reconstruct the accretion history from the definitive RXTE ASM dwell data, using the total count rate and the rates in the three energy bands. Within each observed day the dwells are combined with inverse-variance weights; observed days receive equal weight in each coarse interval. Negative measured rates are retained, and the superburst interval is excluded. The conversion to unabsorbed 0.01--100~keV flux assumes an absorbed cutoff spectrum with $N_{\rm H}=1.3\times10^{22}$~cm$^{-2}$ and cutoff energy 5.2~keV. Its effective slope is determined from the upper two ASM bands, using the instrumental effective area and the persistent flux normalization before the superburst. The resulting spectrum is extrapolated beyond the ASM bands.

\setlength{\skip\footins}{0pt}The persistent bolometric flux is normalized to $10^{-8}$~erg\,cm$^{-2}$\,s$^{-1}$ before the superburst \citep{kuulkers2002}, using the 28 observed days ending two days before the event. The conversion to mass accretion rate assumes isotropic emission and a radiative efficiency of 20\% of the accreted rest-mass energy. Before ASM monitoring, we adopt a constant accretion rate of $10^{17}$~g\,s$^{-1}$ from MJD~47365.5 to 50088.0. For two intervals with poorly constrained ASM band ratios, we interpolate the spectral slope while retaining the measured total rate. Across the final monitoring gap of about 59~d, the flux is interpolated linearly to the last pointed 2--10~keV value, $2.3\times10^{-10}$~erg\,cm$^{-2}$\,s$^{-1}$, with bolometric factor 1.8.

The count-rate errors in Figure~\ref{fig:timeline}a are propagated in quadrature assuming independent dwells. These errors include a systematic component\footnote{\url{https://heasarc.gsfc.nasa.gov/docs/xte/asm_products.html}} but not correlations between dwells. The ASM data extend through September 2011.

The adopted superburst epoch is MJD~50349.42, and cooling times are measured from MJD~51930.5, when accretion ends in the thermal histories. ASM times are converted from TT to UTC; quiescent exposure times retain their arithmetic midpoints in TT, without live-time weighting. We integrate the adopted accretion history from its start to the superburst and divide the accreted mass by $4\pi R^2$. The newly accreted column is approximately $2.1\times10^{12}$~g\,cm$^{-2}$ at 4.5~kpc and $5.1\times10^{12}$~g\,cm$^{-2}$ at 7~kpc. These columns include the assumed constant accretion before monitoring.

The superburst fluxes and error bar lengths are digitized from the published plot. The weighted sums of squared residuals use these errors as scales; their original confidence level and covariance are unknown. The spectra used to derive the burst fluxes share a subtracted persistent emission spectrum \citep{kuulkers2002}, whose uncertainty could correlate different time intervals. These sums describe agreement with the decay, separately from $W$. Appendix~\ref{app:releaseprofile} uses them to construct a likelihood for conditional flux ranges under explicit assumptions about the errors.

The preliminary superburst-decay fits used to select the fuel depths start from a uniform temperature of $2\times10^8$~K. The uniform energy deposited per gram is fitted independently of the carbon instability condition. At each trial depth and energy, we fit the onset without an additional flux multiplier and compare the models using the weighted sum of squared flux residuals. Both a nickel layer with a tabulated envelope and an iron layer with an analytic envelope favor $y=6.3\times10^{11}$~g\,cm$^{-2}$ over the neighboring sampled depth of $10^{12}$~g\,cm$^{-2}$. We use these two depths and their geometric midpoint for the heating comparison.

\section{Superburst decay from the fitted thermal histories}
\label{app:burstdecay}

\subsection{Uniform Release at Adopted Fuel Depths}
\label{app:uniformrelease}
\enlargethispage{-12.5pt}

At the smallest fuel depth, the initial temperatures are those at $t_{\rm sb}$ in the two histories fitted with $X_{\rm C}=0.20$ and local carbon instability (Table~\ref{tab:ksfits}). The initial profiles extend outward to $y\simeq4.20\times10^{10}$~g\,cm$^{-2}$ and are interpolated in $\log y$ and $\log T$. At smaller columns, a steady Fe envelope with an analytic surface boundary extends to approximately $10^8$~g\,cm$^{-2}$ and matches the temperature near the outer end of the fitted thermal profile. The temperature-matched envelope has a flux approximately 18\% higher than that of the fitted accretion history near the joining depth. Each burst evolution is paired with an otherwise identical evolution without burst energy release, with the same envelope, ongoing heating and boundary conditions.

We calculate the decay with a modified version of burstcool,\footnote{\url{https://github.com/andrewcumming/burstcool}} based on the layer cooling calculations of \citet{cumming2004} and \citet{cumming2006}. We use Fe ashes and liquid conductivity. Thermal neutrino losses from electron--positron annihilation, plasmon decay and bremsstrahlung are included throughout the evolution. We evolve the temperature in 150 radial grid zones, with an inner boundary at $y\simeq6.12\times10^{13}$~g\,cm$^{-2}$. The local inner boundary flux is that of the fitted history, with positive values directed outward. The heating profile at $t_{\rm sb}$ is mapped conservatively onto the decay grid and held fixed during the short evolution; the reconstructed accretion rate is constant over this interval.

A uniform burned carbon mass fraction $X_{\rm C}$ supplies a specific energy $E_{\rm dep}=10^{18}X_{\rm C}$~erg\,g$^{-1}$ for an approximate yield of $10^{18}$~erg per gram of carbon burned to iron-group ashes. At column $y$, let $T_i(y)$ and $T_f(y)$ be the local temperatures immediately before and after energy release, respectively, and let $c_{P,{\rm Fe}}(P,T)$ be the Fe specific heat capacity at constant pressure. The final temperature in each heated zone is obtained from
\begin{equation}
 \int_{T_i(y)}^{T_f(y)}c_{P,{\rm Fe}}(P,T)\,dT=E_{\rm dep}.
 \label{eq:burstenergy}
\end{equation}
\par
\pagebreak[3] The temperature profiles with and without burst energy release evolve for $9\times10^4$~s of local proper time. For full-surface isotropic emission, the excess flux is the difference in their observer-frame luminosities divided by $4\pi d^2$. We interpolate this flux linearly in observer time and integrate over each observed exposure. For each of the two accretion-heating backgrounds, we divide each interval-averaged flux residual by the upper plotted error for an overprediction or the lower error for an underprediction, then sum the squared values. We fit the common abundance and onset by minimizing the mean of the two sums, with both histories compared against the same 18 measurements.

We apply this decay calculation at each adopted fuel depth, initially holding the temperature profiles, heating strengths and core temperatures fixed. We vary the shared abundance over $0.10\leq X_{\rm C}\leq0.40$. For all decay fits in this appendix, onset is measured in observer time from the precursor and lies between the precursor and the start of the first exposure. We refine the abundance locally, fitting onset at each trial. The fixed-abundance fit uses $X_{\rm C}=0.20$ and fits only the onset.

For the thermal refits, we evaluate Equation~\ref{eq:ignition} at the fitted abundance using $X_{\rm Fe}=1-X_{\rm C}$. The HZ90 crust and background thermal transport remain unchanged. We recalculate both decays on the refitted thermal histories and refit their common carbon abundance and onset. The burst-heated temperature profile is evolved only through the short decay; quiescent predictions use the accretion-heated histories.

\begin{table}[htbp]
\centering
\caption{Decay and cooling fits with uniform burst energy release.}
\label{tab:depthdecay}
\setlength{\tabcolsep}{7pt}
\begin{tabular}{lrrrrr}
\toprule
Rate & \shortstack{$y_*$\\($10^{11}$ g\,cm$^{-2}$)} & $X_{\rm C}$ & \multicolumn{2}{c}{Spectral Statistic $W$} & \shortstack{Mean Weighted\\Sum of Squared\\Residuals}\\
\cmidrule(lr){4-5}
 & & & Shallower & Deeper & \\
\midrule
CF88 & 6.30 & 0.232 & 303 & 604 & 62.2\\
CF88 & 7.94 & 0.226 & 294 & 561 & 26.6\\
CF88 & 10.00 & 0.215 & 287 & 523 & 59.0\\
T18 & 7.94 & 0.227 & 271 & 435 & 26.4\\
\bottomrule
\end{tabular}
\par\medskip
\begin{tabular}{lrlrrr}
\toprule
Rate & \shortstack{$y_*$\\($10^{11}$ g\,cm$^{-2}$)} & Region & \shortstack{$q_a$\\(MeV/nucleon)} & \shortstack{$T_{\rm c}$\\($10^7$ K)} & \shortstack{$N_{\rm H}$\\($10^{22}$ cm$^{-2}$)}\\
\midrule
CF88 & 6.30 & Shallower & 2.15 & 7.81 & 1.58\\
CF88 & 6.30 & Deeper & 3.93 & 8.27 & 2.44\\
T18 & 7.94 & Shallower & 1.65 & 7.76 & 1.42\\
T18 & 7.94 & Deeper & 3.01 & 8.05 & 2.00\\
\bottomrule
\end{tabular}
\end{table}

For CF88 at the smallest depth, the best evaluated decay fits place the onset at the precursor. This holds for both the original histories fitted with $X_{\rm C}=0.20$ and those refitted using the carbon abundance inferred from the decay. The lower part of Table~\ref{tab:depthdecay} gives the heating and spectral parameters for this comparison and for T18 at the middle depth.

We evolve the decay with adaptive time steps and relative tolerance $10^{-10}$. For CF88 at the fitted abundance, increasing the grid from 150 to 250 zones changes every exposure-averaged flux by less than 0.12\% for both the original and refitted histories. We hold the abundance, onset, physical boundaries, fuel depth, profile joining depth and heating inputs fixed.

For CF88, refitting the abundance on the updated backgrounds changes it by less than 0.001 and the mean weighted sum of squared residuals by less than 0.01 (Table~\ref{tab:depthdecay}).

We also apply this procedure with T18 at the middle fuel depth. The thermal refits follow the T18 search with five core temperatures described in Appendix~\ref{app:search}. Both refitted histories satisfy Equation~\ref{eq:ignition} at the common abundance $X_{\rm C}=0.227$. We evaluate four trial abundances over $0.224\leq X_{\rm C}\leq0.230$ on these updated backgrounds. The fitted abundance is unchanged, with the common onset at the precursor. Updating the thermal backgrounds changes every exposure-averaged decay flux by less than 0.08\%. The two refitted thermal histories give exposure-averaged decay fluxes differing by less than 0.06\%.

\subsection{The Distribution of Burst Energy}
\label{app:releaseprofile}

The fixed CF88 thermal backgrounds are those refitted at $y_*=7.94\times10^{11}$~g\,cm$^{-2}$ and $X_{\rm C}=0.226$. The outer boundary is at $8.74\times10^7$~g\,cm$^{-2}$. The temperature is matched to the outer envelope at $4.88\times10^{10}$~g\,cm$^{-2}$; the heat flux is not constrained to be continuous there.

Let $e_{\rm b}(y)$ be the burst energy deposited per gram at column $y$, and $\bar e_{\rm b}$ its mass-weighted mean over the heated layer, both in erg\,g$^{-1}$. For an exponent $\beta$ fixed at zero or $1/4$, we prescribe $e_{\rm b}(y)\propto(y/y_{\rm h})^\beta$ between the outer boundary and $y_{\rm h}$, with zero deposition below $y_{\rm h}$. The energy is integrated over each radial grid zone, including the heated fraction of the zone containing $y_{\rm h}$, and normalized to the same total for equal $y_{\rm h}$ and $\bar e_{\rm b}$. Replacing $E_{\rm dep}$ by $e_{\rm b}(y)$ in Equation~\ref{eq:burstenergy} determines the temperature increase.

The energy release increasing with depth is motivated by the temperature slope inferred from the early superburst emission of 4U~1636$-$536 \citep{keek2015}. For relativistic degenerate electrons and negligible initial temperature, energy release proportional to $y^{1/4}$ gives approximately $T\propto y^{1/4}$. Here the adopted heat capacity and initial thermal profile determine the final temperature. The mean energy deposited per gram is fitted independently of carbon abundance. Relating it to the local abundance requires a choice of energy yield and burned fraction as functions of depth.

For each background and energy distribution, the heated depth ranges from $3.5\times10^{11}$ to $1.6\times10^{12}$~g\,cm$^{-2}$ and the mean energy deposited per gram from $10^{17}$ to $4\times10^{17}$~erg\,g$^{-1}$. No additional flux multiplier is fitted. We begin with a $5\times5$ grid in heated depth and mean deposited energy, then refine the fit locally three times, optimizing onset for each calculated decay.

For the conditional flux ranges in Figure~\ref{fig:burstdecay}, we integrate jointly over $y_{\rm h}$, $\bar e_{\rm b}$ and onset on the fixed shallower thermal background, separately for each $\beta$. We assume independent flux measurements and interpret the digitized upper and lower errors as standard deviations. The likelihood is proportional to the exponential of minus one-half the weighted sum of squared residuals. Priors are uniform in $\log y_{\rm h}$, $\log\bar e_{\rm b}$ and onset within the bounds above. The shaded ranges enclose the 16th--84th percentiles of the predicted flux at each time. Priors uniform in depth and energy instead shift these flux limits by less than 0.3\%. Treating the digitized errors as 90\% Gaussian intervals narrows the ranges to about 60\% of their nominal width; increasing the error scales by 50\% widens them by approximately the same proportion.

The parameter integration uses fluxes interpolated linearly over triangles in $\log y_{\rm h}$ and $\log\bar e_{\rm b}$, with each curve averaged over the observed exposures before comparison. For twelve parameter choices not used to construct the interpolation, direct calculations give exposure averages that differ from the interpolated values by less than 0.01 times the plotted flux errors.

For the 7~kpc decay comparison, we instead use the thermal backgrounds refitted at that distance with $y_*=6.3\times10^{11}$~g\,cm$^{-2}$ and $X_{\rm C}=0.20$. The upper search limits are extended to $3.4\times10^{12}$~g\,cm$^{-2}$ in heated depth and $8\times10^{17}$~erg\,g$^{-1}$ in mean energy. For release increasing with depth, the onset reaches its upper bound at the start of the first exposure.

The total energy deposited locally by the burst, $\mathcal{E}_{\rm b}$, is the specific deposited energy integrated over the heated mass across the full stellar surface. For a given release distribution, the fitted depths, energies and weighted sums of squared residuals agree between the two backgrounds at the precision shown in Table~\ref{tab:burstfits}. At 4.5~kpc, the uniform-release fits place onset at the precursor, so their entries also apply when onset is fixed there.

\begin{table}[!htbp]
\centering
\caption{Selected decay fits on the fixed thermal backgrounds at each distance.}
\label{tab:burstfits}
\setlength{\tabcolsep}{5pt}
\renewcommand{\arraystretch}{1.12}
\begin{tabular}{rrrrrrr}
\toprule
$\beta$ & \shortstack{$y_{\rm h}$\\($10^{11}$ g\,cm$^{-2}$)} & \shortstack{$\bar e_{\rm b}$\\($10^{17}$ erg\,g$^{-1}$)} & \multicolumn{2}{c}{Onset (s)} & \shortstack{$\mathcal{E}_{\rm b}$\\($10^{42}$ erg)} & \shortstack{Weighted Sum\\of Squared\\Residuals}\\
\cmidrule(lr){4-5}
 & & & Shallower & Deeper & & \\
\midrule
\multicolumn{7}{l}{4.5~kpc, onset fitted}\\
0 & 8.00 & 2.26 & 0.0 & 0.0 & 2.27 & 26.6\\
$1/4$ & 6.62 & 2.63 & 125.9 & 126.9 & 2.19 & 10.2\\
\addlinespace[4pt]
\multicolumn{7}{l}{4.5~kpc, onset fixed at the precursor}\\
$1/4$ & 6.79 & 2.67 & 0.0 & 0.0 & 2.27 & 10.7\\
\addlinespace[4pt]
\multicolumn{7}{l}{7~kpc, onset fitted}\\
0 & 17.1 & 3.83 & 594.6 & 594.7 & 8.22 & 321\\
$1/4$ & 10.3 & 5.25 & 968.7 & 968.7 & 6.77 & 210\\
\bottomrule
\end{tabular}
\end{table}

Increasing the radial resolution from 150 to 250 zones changes the predicted exposure averages by less than 0.25\% for all four selected 4.5~kpc fits with fitted onset. At 7~kpc, the corresponding changes remain below 1\% for the selected uniform-release fit and a neighboring model in which energy release increases with depth ($\bar e_{\rm b}=5.0\times10^{17}$~erg\,g$^{-1}$). In both cases, the shallower thermal background and burst parameters are held fixed as the grid is refined.

\begin{figure*}[t]
\centering
\figfile[width=0.88\linewidth]{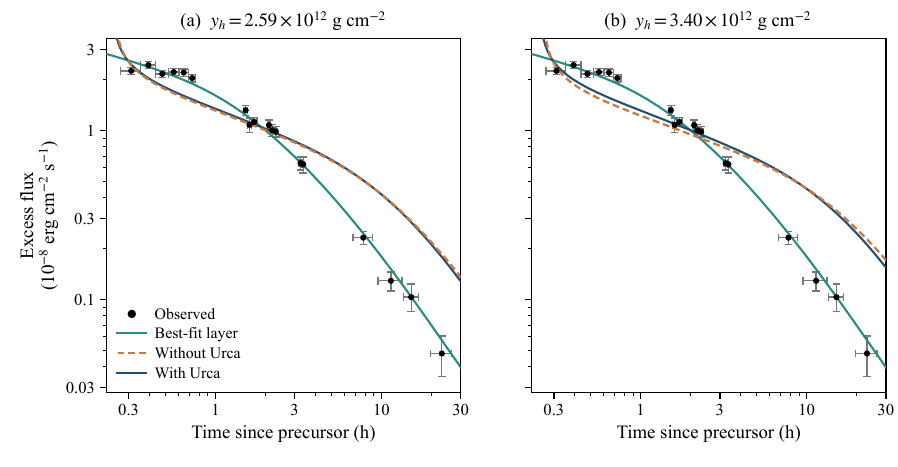}
\caption{Superburst decays for uniform energy release extending to the Urca shell (a) and below it (b), on the fixed thermal background from the shallower heating region at 4.5~kpc. Curves show models with and without $^{23}$Na--$^{23}$Ne cooling, fitted to the exposure-averaged fluxes; the best-fitting uniform-release model has $y_{\rm h}=8.0\times10^{11}$~g\,cm$^{-2}$. Points show the fluxes from \citet{kuulkers2002}, with gray vertical bars for the digitized errors and horizontal bars for the observing intervals.}
\label{fig:urca}
\end{figure*}

\subsection{Burst Cooling at the Sodium--Neon Urca Shell}
\label{app:urca}

The $^{23}$Na--$^{23}$Ne pair has the largest cooling strength per unit mass fraction among the ocean Urca pairs tabulated by \citet{deibel2016}. Its electron-capture threshold of 4.38~MeV corresponds to $P\simeq6.3\times10^{26}$~dyn\,cm$^{-2}$, or $y\simeq2.6\times10^{12}$~g\,cm$^{-2}$ for the adopted gravity. We use their thin-shell luminosity with a dimensionless nuclear strength of 17 and a prescribed total pair mass fraction of 0.067, held fixed throughout the decay. The luminosity scales with the fifth power of local temperature and enters the thermal energy equation.

On the two 4.5~kpc thermal backgrounds from Appendix~\ref{app:releaseprofile}, we extend the burst-heated layer to the capture depth or to $3.4\times10^{12}$~g\,cm$^{-2}$. At each depth and for $\beta=0$ and $1/4$, we fit onset and mean energy independently for the decays with and without Urca cooling, allowing $2\times10^{16}\leq\bar e_{\rm b}\leq6\times10^{17}$~erg\,g$^{-1}$. The paired evolution without burst energy release uses the same Urca cooling prescription. We distribute the shell loss between the two zones bracketing the capture pressure, using fixed weights linear in $\log P$ that sum to unity. Each zone's contribution is evaluated at its evolving temperature. Increasing the grid from 150 to 250 zones changes the exposure-averaged fluxes of the fitted models by less than 0.14\%.

For the models in Figure~\ref{fig:urca}, we integrate over $\bar e_{\rm b}$ and onset at fixed heated depth, thermal background and Urca strength. We use the likelihood and error assumptions of Appendix~\ref{app:releaseprofile}, with priors uniform in $\log\bar e_{\rm b}$ and onset within the bounds above.

For uniform release at either tested depth, the conditional 16th--84th percentile interval for the mean predicted-to-observed flux ratio over the final four exposures lies close to three. Energy release increasing with depth produces a larger late-time flux excess, and changing the thermal background has little effect on the fitted decays. The additional neutrino loss relative to the evolution without burst energy release accounts for about 5--10\% of the deposited energy over the first 32.5 observer-frame hours; both energies are evaluated locally. At the shallower heated depth $y_{\rm h}=8.0\times10^{11}$~g\,cm$^{-2}$, adding the same Urca cooling to the uniform-release fit changes the predicted flux by less than 0.07\%.

We test whether concentrating the Urca loss in two zones affects the late excess. For uniform release to the capture depth on the shallower thermal background, we distribute the loss with Gaussian weights in electron chemical potential centered at 4.38~MeV. The fixed standard deviation, $0.37$~MeV, equals $k_B T$ at the shell's peak temperature without Urca cooling; the weights preserve the same luminosity at uniform temperature. After locally refitting $\bar e_{\rm b}$ and onset, the late excess remains of the same magnitude.

\raggedbottom
\clearpage\apptocmd{\thebibliography}{\setlength{\baselineskip}{12.5pt}\setlength{\bibskip}{0pt}}{}{\PackageError{aastex-layout}{Bibliography spacing patch failed}{}}

\end{document}